\documentclass[fleqn,usenatbib]{mnras}

\usepackage{graphicx}	
\usepackage{amsmath}	
\usepackage{amssymb}	
\usepackage{tabularx}
\usepackage[flushleft]{threeparttable}
\usepackage[a4paper]{geometry}%
\usepackage{hyperref}

\newcommand{\Msun}{~M_\odot}

\newcommand{\cmc}{\rm ~cm^{-3}}

\newcommand{\ergs}{\rm ~erg~s^{-1}}

\title[Sound wave generation by an outburst]{Sound wave generation by a spherically symmetric outburst and AGN Feedback in Galaxy Clusters}

\author[Xiaping Tang and Eugene Churazov]{
Xiaping Tang,$^{1}$\thanks{E-mail: xt5uv@mpa-garching.mpg.de}
Eugene Churazov,$^{1,2}$
\\
$^{1}$Max Planck Institute for Astrophysics,
Karl-Schwarzschild-Str. 1,
D-85741 Garching, Germany\\
$^2$ Space Research Institute (IKI), Profsoyuznaya 84/32, Moscow 117997, Russia
}

\date{Accepted XXX. Received YYY; in original form ZZZ}

\pubyear{2017}

\begin{document}
\label{firstpage}
\pagerange{\pageref{firstpage}--\pageref{lastpage}}
\maketitle

\begin{abstract}
We consider the evolution of an outburst in a uniform medium under spherical symmetry, having in mind AGN feedback in the intra cluster medium (ICM). For a given density and pressure of the medium, the spatial structure and energy partition at a given time $t_{age}$ (since the onset of the outburst) are fully determined by the total injected energy $E_{inj}$ and the duration $t_b$ of the outburst. We are particularly interested in the late phase evolution when the strong shock transforms into a sound wave. We studied the energy partition during such transition with different combinations of $E_{inj}$ and $t_b$. For an instantaneous outburst with $t_b\rightarrow 0$, which corresponds to the extension of classic Sedov-Taylor solution with counter-pressure, the fraction of energy that can be carried away by sound waves is $\lesssim$12\% of $E_{inj}$. As $t_b$ increases, the solution approaches the ‘'slow piston'' limit, with the fraction of energy in sound waves approaching zero. We then repeat the simulations using radial density and temperature profiles measured in Perseus and M87/Virgo clusters. We find that the results with a uniform medium broadly reproduce an outburst in more realistic conditions once proper scaling is applied.  We also develop techniques to map intrinsic properties of an outburst $(E_{inj}, t_b$ and $t_{age})$ to the observables like the Mach number of the shock and radii of the shock and ejecta.  For the Perseus cluster and M87, the estimated $(E_{inj}, t_b$ and $t_{age})$ agree with numerical simulations tailored for these objects with $20-30\%$ accuracy.
\end{abstract}


\section{Introduction}
\label{sec:intro}

The problem of a powerful outburst in a nearly
homogeneous medium under spherical symmetry plays an important role in a number of astrophysical
contexts, ranging from supernova remnants to AGN feedback in the cores
of galaxy clusters. In the latter case, particularly important
questions are (i) deriving the energetics of the outburst from 
readily available observables and (ii) determining what fraction of
the outburst energy is going to be thermalized in the gas. They are
also the questions that we address in this paper.

The importance of AGN activity on the ICM thermal state in clusters
and elliptical galaxies is nowadays widely accepted
\citep[e.g.,][]{2000A&A...356..788C,2000ApJ...534L.135M}, see e.g., \citet{2012ARA&A..50..455F} and \citet{Soker16} for reviews. The cooling time of the ICM in cluster cores is
short and a source of energy is needed to prevent a massive build up
of cold gas and vigorous star formation. This energy is believed to
come from a central supermassive black hole, although the description
of the energy transfer from AGN to ICM is still
sketchy. Multi-wavelength observations reveal that AGN jets inflate
bubbles of relativistic plasma (or at least a mixture of relativistic particles and very hot plasma). When energy
is injected in an unperturbed medium, the expanding bubble initially
drives a strong shock in the surrounding medium and heats the
gas. Later, the bubble expansion becomes subsonic and the shock-heating efficiency drops dramatically. The energy
goes instead into the thermal energy of the expanding bubble and $PV$
work. Eventually, bubbles are removed from the cluster core by the
buoyancy force. During the process, the rising bubbles must loose their energy after crossing several pressure scale heights \citep{2001ApJ...554..261C,2002MNRAS.332..729C,2001ASPC..250..443B} and
can excite turbulence
and/or internal waves that are eventually dissipated into heat \citep[e.g.,][]{2014Natur.515...85Z}.  Sound
waves could also contribute to the gas heating \citep[e.g.,][]{2006MNRAS.366..417F}, if a substantial fraction
of energy goes into the waves and they are able to dissipate. Yet another scenario postulates that the
bubbles are filled with hot, but non-relativistic, plasma and mixing
of this plasma with the ICM is the dominant channel for ICM heating \citep[e.g.,][]{2016MNRAS.455.2139H}.

Which of these scenarios dominate depends on the way AGN injects energy
to the ICM. While this problem has been addressed in many
sophisticated 3D numerical simulations, the uncertainties in choosing
the right approximation and parameters precludes unambiguous conclusion. A simpler approach is to use the minimalistic set of
assumptions  to get order of magnitude constraints on a few of the most important
parameters. For instance, one-dimensional (1D) spherically symmetric outburst powered by
AGN activity is probably the simplest among all models. It was
studied, e.g., in \citet{Forman07,Forman16,Graham08,2011ApJ...726...86R,Zhuravleva16} where constraints on the energy/luminosity
and duration of the outburst have been obtained. In all these cases
the simulations were tailored for a particular object. Several
questions arise: (i) are the results obtained for well studied systems
universal, (ii) what are the most extreme cases allowed by models,
e.g., maximal amount of energy in the form of sound waves, (iii) can one
use a few observables to estimate the properties of an outburst without
doing actual simulations? To address these question we made a
systematic study of an outburst evolution in a very simple setup and
derive a scaling needed to apply this analysis to other systems.
  
In this paper we closely investigate the evolution of an outburst in a low density hot medium under spherical symmetry. Qualitatively, ``low
density'' is assumed to indicate that the radiative energy losses are
negligible on time scales of the outburst evolution. The assumption of
``hot'' means that despite  the low density, the thermal pressure in
the ambient medium is not negligible and is dynamically important for
the expansion of an outburst. Because of the high pressure in the
medium, the shock wave driven by the outburst gradually transforms
from a strong shock into a weak shock and eventually asymptotically approaches a
sound wave as $t\rightarrow \infty$.

In \S\ref{sec:simple}, we describe the main assumptions/approximation
made in the paper. In \S \ref{uniform medium}, we systematically study the
evolution of an outburst in uniform medium. In
\S \ref{non-uniform}, we use the Perseus cluster and M87 as examples to show that the
evolution of an outburst in a realistic density and pressure profile of a
galaxy cluster does not deviate much from those results in uniform medium. We then develop simple tools to map the observables to the intrinsic properties of an outburst based on simulation with uniform medium. In \S \ref{sec:classification}, we create a schematic figure to illustrate the various regimes and physical processes related to the problem of AGN driven outburst and then discuss the physical properties of the outbursts in Perseus cluster and M87. \S \ref{sec:conclusion} is the conclusion section.

\section{Outline of the problem}
\label{sec:simple}
The primary goal of this paper is to use 1D
spherically symmetric outburst model to derive the most basic constraints
on the spatial structure and energy partition of AGN driven outbursts in galaxy cluster. Here we outline main assumptions and simplifications made in our discussion.

The initial gas density $\rho_a(r)$ and pressure $P_a(r)$ distributions of the ICM are assumed to be static, where $r$ is the distance from the cluster center. In the presence of pressure
gradients, a balancing force due to gravity is taken into account to maintain hydrostatic equilibrium. Thermal conduction and physical viscosity are neglected for simplification. At age $t_{age}=0$, energy and mass start to be injected into the center of the cluster. In our simplified problem, for a given $\rho_a(r)$ and $P_a(r)$ profiles of the ICM, the evolution of an outburst is fully determined by its energy and mass injection history. In this paper, we consider simplified energy and mass injection (see Appendix \ref{ap:a} for a detailed description) which are characterized by three parameters: total injected energy $E_{inj}$, total injected mass $M_{inj}$ and the duration of the outburst $t_b$. The adiabatic index of the injected material is
$\gamma_{ej}$, while the ambient gas has the adiabatic index $\gamma_a$. In the context of galaxy clusters, this setup is very similar to the one used in, e.g.,
\citet{1998ApJ...501..126H,Forman07,Zhuravleva16}.

The fate of such outbursts could fall into two different evolution tracks depending on the value of the dimensionless ratio $(E_{inj}/M_{inj})/(P_a/\rho_a)$, see Appendix \ref{ap:b0} for detailed discussion. In most astrophysical applications, like the AGN feedback problem studied here, the ratio $(E_{inj}/M_{inj})/(P_a/\rho_a)\gg 1$ and the evolution of the outburst could be divided into three different phases, if we make analogy to instantaneous outburst with $t_b\rightarrow0$ for outbursts with finite duration time $t_b$. Based on the relation between the injected energy $E_{inj}$ and mass $M_{inj}$ and the swept up materials thermal energy $E_{sw}$ and mass $M_{sw}$, the three phases are free expansion phase with $M_{sw}\leq M_{inj}$ and $E_{sw}\ll E_{inj}$, Sedov-Taylor (ST) phase \citep{Taylor46,Sedov59} with $M_{sw}> M_{inj}$ and $E_{sw}\leq E_{inj}$, and a wave-like phase with $E_{sw}> E_{inj}$. For the AGN feedback problem of interest here, $M_{inj}$ is unknown from observation but likely to be small, and the free expansion phase is likely to be very brief. So in the rest of the paper, we could simply neglect the injected mass $M_{inj}$ and ignore the free expansion phase which are not very important for our discussion. In view of the weak shock observed in galaxy cluster, we will focus on the transition from ST phase to wave-like phase in this paper. 

The transition from ST phase to the wave-like phase happens when the thermal energy of swept up materials $E_{sw}$ becomes comparable with the injected energy $E_{inj}$. It sets the characteristic length $R_E$ and time $t_E$ of the system, see Appendix \ref{ap:b} for detail. $R_E$ is the characteristic radius of a sphere such that the total
thermal energy of the ICM within the sphere is of order of $E_{inj}$, i.e.
\begin{equation}
E_{inj}=\frac{4\pi}{3}P_a R_E^3, 
\label{eq:charac_R}
\end{equation}
where we have assumed for simplicity that the pressure $P_a(r)={\rm
  const}$. $t_E$ is the sound crossing time of $R_E$, i.e.
\begin{equation}
t_E=\frac{R_E}{\sqrt{\gamma_a P_a/\rho_a}}. 
\label{eq:charac_t}
\end{equation}
In the subsequent sections, we will study the evolution of an outburst during such transition for
different combinations of $t_b$ and $t_E$. Two special cases that deserve more discussions are (i) the instantaneous outburst with $t_b/t_E\rightarrow 0$ and (ii) the continuous outburst (or
``slow piston'' mode) with $t_b/t_E\rightarrow \infty$.  They behave very differently during the transition from the early ST phase to the late wave-like phase especially in energy partition. In the first case, a large amount of energy is released during a short period of time. When the energy release
quenches, a strong shock still exists which could heat the surrounding ambient gas and generate a significant amount of entropy. We also expect that the state of such an outburst at age $t_{age}\gg t_b$ will
not be very sensitive to the actual value of $t_b$. In the second case, the injected energy is released over a time scale much larger than the characteristic time $t_E$ of the system. As a result, the ejecta always expand subsonically, except for a short interval at early time, and no net entropy generation in the ICM is happening.

Extensive efforts have been made for the two limiting cases by previous analytical work. For an instantaneous outburst, it follows the classic ST solution at $t_{age}/t_E\rightarrow 0$ \citep{Taylor46,Sedov59} while at $t_{age}/t_E\rightarrow \infty$ the shock wave driven by the outburst asymptotically approaches a sound wave due to the ambient pressure. The transition from classic ST solution to the asymptotic wave-like solution is studied by \cite{Melnikova54} and \cite{Sakura54} according to perturbation theory but only valid when the deviation from ST solution is not very large. The asymptotical behavior of the outburst at $t_{age}/t_E\rightarrow \infty$ has been studied by \cite{Landau45} and \cite{Bethe58} but focusing on the decay of the shock front. The global structure and energy partition of the outburst however are not discussed in detail. For a continuous outburst, the self-similar solution as an extension of the classic ST solution is available at $t_{age}/t_E\rightarrow 0$ \citep{Dokuchaev02,2016PASJ...68...22M}, if the energy injection in space and time satisfies certain conditions. At $t_{age}/t_E\rightarrow \infty$, the shock front gradually decays due to ambient pressure and eventually disappear. No wave-like structure is generated in a continuous outburst, because the energy injection never quenches and thus no rarefaction structure is formed behind the shock front. 

The transition between the asymptotic ST solution at $t_{age}/t_E\rightarrow 0$ to the asymptotic wave-like solution at $t_{age}/t_E\rightarrow \infty$ is difficult to model analytically and will be studied numerically. The time evolution of an outburst with finite $t_b$ hasn't been discussed before and will become the focus of this paper. Our final goal is to link the observables to all major characteristics of the
outburst, including $E_{inj}$, $t_b$ and $t_{age}$, and then differentiate between cases according to different $t_b/t_E$ ratios. Among these observables the
most easily accessible are: the Mach number $M$ and the radius of the
shock front $R_s$, and the radius of the ejecta $R_{ej}$. The latter quantity can be derived from observations by measuring the size of ``radio bubbles'' inflated by an AGN. 
We therefore investigate
the relation between these observables and the characteristic properties of the outburst in detail for different combinations of $E_{inj}$, $t_b$ and $t_{age}$.  In section \ref{uniform medium}, we present the results for uniform ambient medium. Next, in section \ref{non-uniform},
we discuss how a more realistic radial profile of the ambient medium affects
the results from uniform medium case. The numerical method used in this paper
is described in Appendix~\ref{ap:num}.

\section{Outburst in a uniform medium}{\label{uniform medium}}
\label{sec:uniform}
In this section, we investigate the dynamical evolution of an outburst in a uniform media with constant density and pressure profile, i.e. $\rho_a(r)=\rho_0$ and $P_a(r)=P_0$. We will focus on the spatial structure and energy distribution of the outburst during the transition from the ST phase to the wave-like phase. The natural way to illustrate the complete physical picture is to divide the problem into different regimes based on the $t_b/t_E$ ratio and then investigate the physical properties of an outburst as a function of $t_{age}/t_E$ in each individual regime closely. In this section, all the results are presented in dimensionless form to keep the generality of our discussion.

In order to demonstrate the energy distribution of an outburst over radius, we calculate cumulative kinetic $E_{kin}$ and thermal $E_{th}$ energies, which are defined as follows
\begin{equation}
E_{kin}(r,t_{age})=\int_0^{r} \frac{1}{2} v^2(r',t_{age})dM(r'),
\end{equation}
where $v(r,t_{age})$ is the flow velocity at radius $r$ and age $t_{age}$,
and 
\begin{equation}
E_{th}(r,t_{age})=\int_0^{r} e(r',t_{age})dM(r')-\int_0^{r} e(r',0)dM(r'),
\label{eq:eth}
\end{equation}
where $M(r')$ is the mass within radius $r'$ and $e(r,0)$ is the initial distribution of internal energy per unit mass in the ambient medium. In the calculation of $E_{th}$, we subtract the initial internal energy in the swept up materials to focus on only the spatial distribution of the injected outburst energy $E_{inj}$. Based on energy conservation $E_{th}(R_s,t)+E_{kin}(R_s,t)=E_{inj}$, where $R_s$ is the radius of an outburst. When $t_{age}/t_E\gg 1$ (or equivalently $R_s(t_{age})/R_E\gg 1$), the initial thermal energy within $R_s$  becomes much larger than the injected outburst energy, i.e. $\int^{R_s}_0 e(r,0)dM(r)\gg E_{inj}$. This limits the accuracy of evaluating $E_{th}$ from eq.~(\ref{eq:eth}) in numerical simulations. Therefore, we constrain all the numerical runs to $t_{age}\leq 10t_E$ to ensure accurate determination of $E_{th}$. 

 \begin{figure}
 \begin{center}
 \includegraphics[width=\columnwidth]{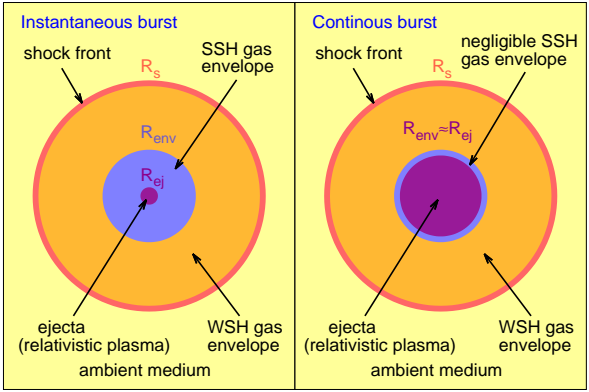} 
  \caption{Sketch  of the spatial structure of an outburst after the shock detachment (see also Forman et al., 2016). The left panel is for an instantaneous outburst and the right panel is for a continuous outburst. SSH and WSH stand for strongly shock heated and weakly shock  heated regions, respectively. $R_s$ is the shock front radius, $R_{env}$ is the boundary that separates the SSH gas envelope from the WSH gas envelope, and $R_{ej}$ is the ejecta radius.} 
    \label{schematic}
 \end{center}
 \end{figure}

We will start with a discussion about the two limiting cases, namely, the instantaneous outburst with $t_b/t_E\rightarrow 0$ and the continuous outburst with $t_b/t_E\rightarrow\infty$. A sketch of the spatial structure for both instantaneous and continuous outbursts at late time $t_{age}\gtrsim t_{E}$ is shown in Fig.1. Here late time refers to the stage that the shock front is moving much faster than the ejecta front, i.e., the shock has detached from the central ejecta bubble.   
In both cases, the affected volume could be divided into three different regions, ejecta (injected hot plasma), strongly shock-heated (SSH) and weakly shock-heated (WSH) gas envelopes, according to their composition and the entropy change during the evolution. In the continuous outburst, the amount of SSH gas is negligible. The physical properties of each region and the explanation of shock detachment process will be discussed in detail in the rest of this section.

\subsection{Instantaneous outburst ($t_b/t_E\rightarrow 0$)}{\label{sec:fastburst}}
The transition from the ST phase to the wave-like phase for an instantaneous outburst has been examined in \cite{T&W05} (see also reference therein) in the context of a supernova explosion with a focus on the kinematics of the shock front. For AGN driven outburst in galaxy or cluster interested here, we instead concentrate on the spatial structure and energy distribution of the outburst. 

\begin{figure}
\begin{center}
\includegraphics[width=\columnwidth]{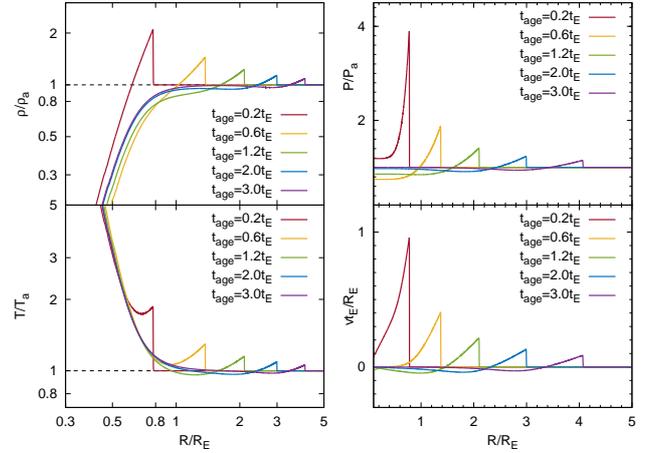} 
\caption{ Radial profiles of $\rho$, $T$, $P$ and $v$ for an instantaneous outburst at different values of $t_{age}/t_E$. All quantities are  normalized to the unperturbed values. The panels for density $\rho$ and temperature $T$ are in log scale while panels for pressure  $P$ and velocity $v$ are in linear scale to emphasize various features emerging during the evolution. Dashed horizontal lines correspond to the unperturbed profiles.} 
    \label{various_age}
\end{center}
\end{figure}

As discussed by \citet{T&W05}, after the ST phase the blast wave driven by the outburst gradually transforms from a strong shock to a weak shock since the internal energy in the swept up materials becomes dynamically important. In Fig. \ref{various_age}, we present the density $\rho$, temperature $T$, pressure $P$ and velocity $v$ profiles for an instantaneous outburst at different values of $t_{age}/t_E$ to illustrate the development of two important features, namely, a shock detachment process and an emergence of a wave-like structure. As $t_{age}/t_E$ increases, a weak shock front gradually detaches from the central bubble with high-temperature and low-density, which slows down with time and eventually becomes static. It is shown in Fig. \ref{various_age} that, especially the panels for temperature and density profiles,  the shock front is expanding away from the central hot bubble with time which implies the existence of a shock detachment process. 
After the shock detachment process, a wave-like structure with both crest and trough in the radial profile starts to emerge behind the shock front, which is more easily seen in the panels for pressure and velocity distribution. During the expansion, the shock velocity asymptotically approaches the speed of sound in the surrounding medium. See Appendix \ref{App:shock_evolution} for a simple analytical approximation of shock evolution in an instantaneous outburst, which is consistent with numerical simulation within a few percent accuracy.

\begin{figure*}
\centering
\includegraphics[width=0.9\textwidth, height=12cm]{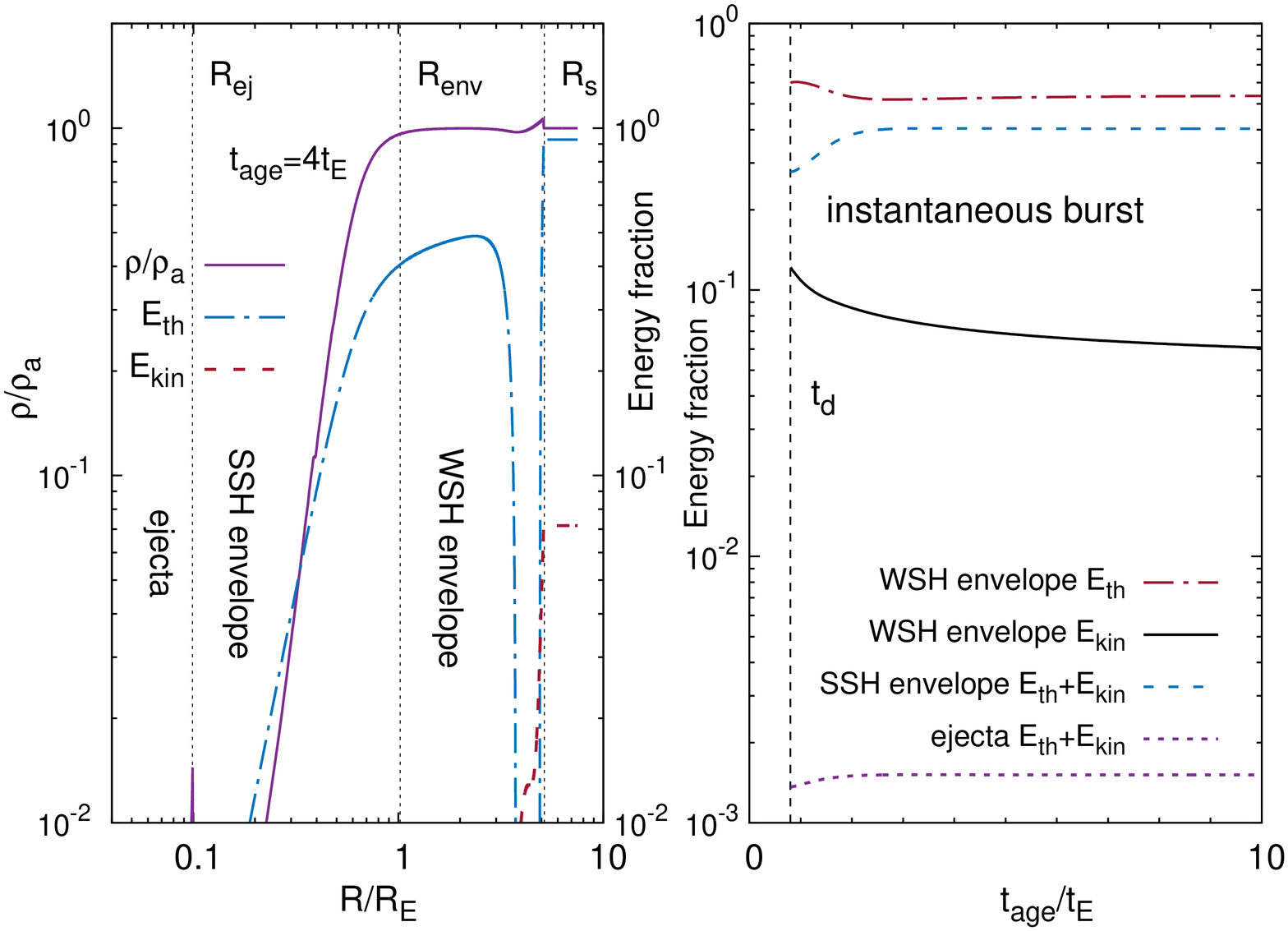}
\caption{Left panel: Spatial structure of an instantaneous
    outburst at $t_{age}=4t_E$. A large fraction of injected energy
    $E_{inj}$ is stored in the SSH envelope while the energy left in
    the ejecta is negligible. A WLS with both crest and trough
    along the radius is presented in the density $\rho$ distribution
    behind the shock front. The thermal energy $E_{th}$ distribution
    shows a sharp decrease and subsequent increase in WLS. Right
    panel: Time evolution of various energy components from shock
    detachment $t_{age}=t_d\approx 0.8t_E$ to $t_{age}=10t_E$. After
    the shock detachment, all energy components quickly approach the
    asymptotic values and then remain almost unchanged.}
\label{various_region}
\end{figure*}

As shown in Fig. \ref{schematic}, the spatial structure of an instantaneous outburst after shock detachment process could be divided into three different regions:  ejecta, SSH and WSH gas envelopes \citep[see, e.g., discussion in][]{Xiang09,Zhuravleva16,Forman16}. In the left panel of Fig. \ref{various_region}, we plot the density $\rho$, cumulative thermal energy $E_{th}$ and kinetic energy $E_{kin}$ distributions of an instantaneous outburst at $t_{age}/t_E=4$ for all three different regions. The physical properties of the three regions are outlined below.

(1). Ejecta (hot plasma injected by an AGN). This is the innermost region with $r<R_{ej}$. 
It has the highest temperature (energy per unit mass) and the lowest density among all three different regions. For an instantaneous outburst, the thermal energy left in the ejecta after shock detachment is negligible as shown in Fig. \ref{various_region}. The kinetic energy of the ejecta is also negligible since at $t_{age}>t_E$ the pressure gradients in the injected hot plasma are quickly erased and the region becomes static.  In observations, the relativistic particles in this region could produce synchrotron radio emission. In X-ray, it is revealed as a cavity/depression in the surface brightness due to the low density/emissivity. 

(2). SSH envelope. This region corresponds to the gas that went through a strong shock and undergoes significant entropy increase. It starts at the outer boundary of ejecta $R_{ej}$ and ends at $R_{env}$. $R_{env}$ is defined as the radius where the entropy change of the shocked medium satisfies $S[r(t_{age}),t_{age}]-S[r(0),0]=0.1(\gamma_a-1)c_v$. $c_v$ is the heat capacity per unit mass. Note $r(t_{age})$ is in Lagrangian coordinate and the definition of $R_{env}$ here is rather arbitrary. In the SSH  region the gas properties gradually change from the hot and low density gas in the inner part to almost unperturbed values at $R_{env}$ where SSH joins the WSH envelope,  as shown in Fig. \ref{various_region}. No prominent synchrotron emission is expected from the SSH region, unless the strong shock leads to particle acceleration. For an instantaneous outburst, the size and amount of gas in this region is significant. So it might be detectable in X-ray observations as a hot and low surface brightness region, enveloping the ejecta. For a continuous outburst, the entropy increase is negligible; thus no SSH region is expected.

(3). WSH envelope. This region between $R_{env}$ and $R_s$ contains
the gas that went through the shock, but the shock was too weak to
induce significant entropy change in the gas.  The WSH region could be
further subdivided into two parts based on their physical
properties. Behind the shock front there is a wave-like structure
(WLS) with both crest and trough along the radius. The WLS starts roughly at the position $R_w$ where $E_{th}$ starts
to decrease with radius, i.e. the smallest radius satisfy
$dE_{th}(r,t)/dr<0$. Between $R_{env}$ and $R_w$, the shocked medium
has physical properties close to the unperturbed value. The two parts
could be easily identified in the $E_{th}$ distribution shown in
Fig. \ref{various_region}. Between $R_{env}$ and $R_w$ the value of
$E_{th}$ increases very slowly, while between $R_w$ and $R_s$, i.e.
in the WLS region, $E_{th}$ first drops steeply and then rise quickly with
radius.  The WLS moves at a speed asymptotically approaching the sound
speed as $t_{age}/t_E\rightarrow \infty$ and carries almost all the
kinetic energy induced by the outburst.

The late time evolution of an instantaneous outburst after the shock detachment is not discussed in detail in \cite{T&W05}. Here we extend their analysis and show the behaviour of different energy components in all three regions after the shock detachment in the right panel of Fig. \ref{various_region}. The shock detachment time $t_d$ is defined in the simulation as the time when a trough shape like structure emerges in $E_{th}$, i.e. the smallest time that satisfy $dE_{th}(r,t)/dr<0$. For an instantaneous outburst, we find $t_d\approx 0.8t_E$ which corresponds to a Mach number $M\approx 1.25$ according to eq (\ref{velocity_fast}) in Appendix \ref{App:shock_evolution}. After the shock detachment, the spatial structure and energy distribution in the ejecta and SSH gas envelope are essentially independent of time as shown in Fig. \ref{various_age} and \ref{various_region}, respectively. At $t_{age}/t_E\gg 1$, the WLS structure carries all the kinetic energy of the system. $E_{kin}$ asymptotically approaches a value $\lesssim 6\%$ of the injected outburst energy $E_{inj}$. As  $t_{age}/t_E\rightarrow\infty$ the WLS becomes a sound wave, and it is expected that the total energy eventually carried away by the WLS would be $E_{th}+E_{kin}=2E_{kin}\approx 12\%$. It provides an estimate of the amount of energy that will eventually leave the system with sound waves.  

\subsection{Slow piston driven outburst ($t_b/t_E\rightarrow \infty$) }{\label{sec: slowburst}}
In the previous section, the energy is released instantaneously and the ejecta are expanding supersonically which drive a strong shock into the ambient gas. We now consider an opposite limit, the so called ``slow piston driven outburst'', in which the ejecta expand very slowly. In this limit, the energy is assumed to be injected at a infinitely slow rate and the ejecta expand extremely slowly with negligible kinetic energy. Under such assumption, no strong shock is generated and no energy dissipation happens in the ICM during the expansion of ejecta. All the injected energy is transferred to the thermal energy of the ejecta and the $PdV$ work, i.e $E_{inj}=PV+PV/(\gamma_{ej} -1)=\gamma_{ej} PV/(\gamma_{ej} -1)$ where $P$ and $V$ are pressure and volume of the inflated ejecta bubble and $\gamma_{ej}$ is the specific heat index of the injected plasma.  For $\gamma_{ej}=5/3$, the total energy, i.e. thermal energy, stored in the ejecta becomes $E_{ej}=E_{inj}/\gamma_{ej}=0.6E_{inj}$ while the remaining $0.4E_{inj}$ is distributed over the swept up ambient medium.

This slow piston driven outburst is an idealized limit. A more plausible case to consider is when the energy is injected at a constant rate $L$. In this case, the ejecta initially expand supersonically, but with time the expansion velocity of ejecta gradually becomes subsonic despite ongoing energy injection. Energy injected during the subsonic phase is accompanied by negligible dissipation at the shock front. As a result, while the total injected energy $E_{inj}=Lt_{age}$ increases linearly with time, the total energy dissipated during the outburst remains almost the same, making the percentage of various energy components in such a continuous outburst resemble the values in a slow piston limit. It is convenient to introduce a new set of characteristic time $t_L$ and length $R_L$, similar to $t_E$ and $R_E$, for a continuous outburst. Basically, in eqs.~(\ref{eq:charac_R}) and (\ref{eq:charac_t}) one sets $E_{inj}=Lt_L$ and replaces $R_E$ and $t_E$ with $R_L$ and $t_L$ (see Appendix \ref{ap:b} for detail discussion). By definition, it is obvious that  $t_L=t_E^{3/2}/t_b^{1/2}$. The time evolution of Mach number $M$ for both an instantaneous outburst and a quasi-continuous outburst with $t_b=10t_E$ is presented in Fig. \ref{Machnumber}. From this Figure, it is clear that the value of M-1 drops below 0.1 at $t_{age}\sim 0.5t_E$ for the quasi-continuous outburst with $t_b=10t_E$, which indicate the outburst has a very weak shock for almost $95\%$ of the outburst duration $t_b$. As $t_{b}/t_E\rightarrow \infty$, the quasi-continuous outburst approaches a continuous outburst with no strong shocks being generated.

\begin{figure*}
 \begin{center}
 \includegraphics[width=0.9\textwidth,height=12cm]{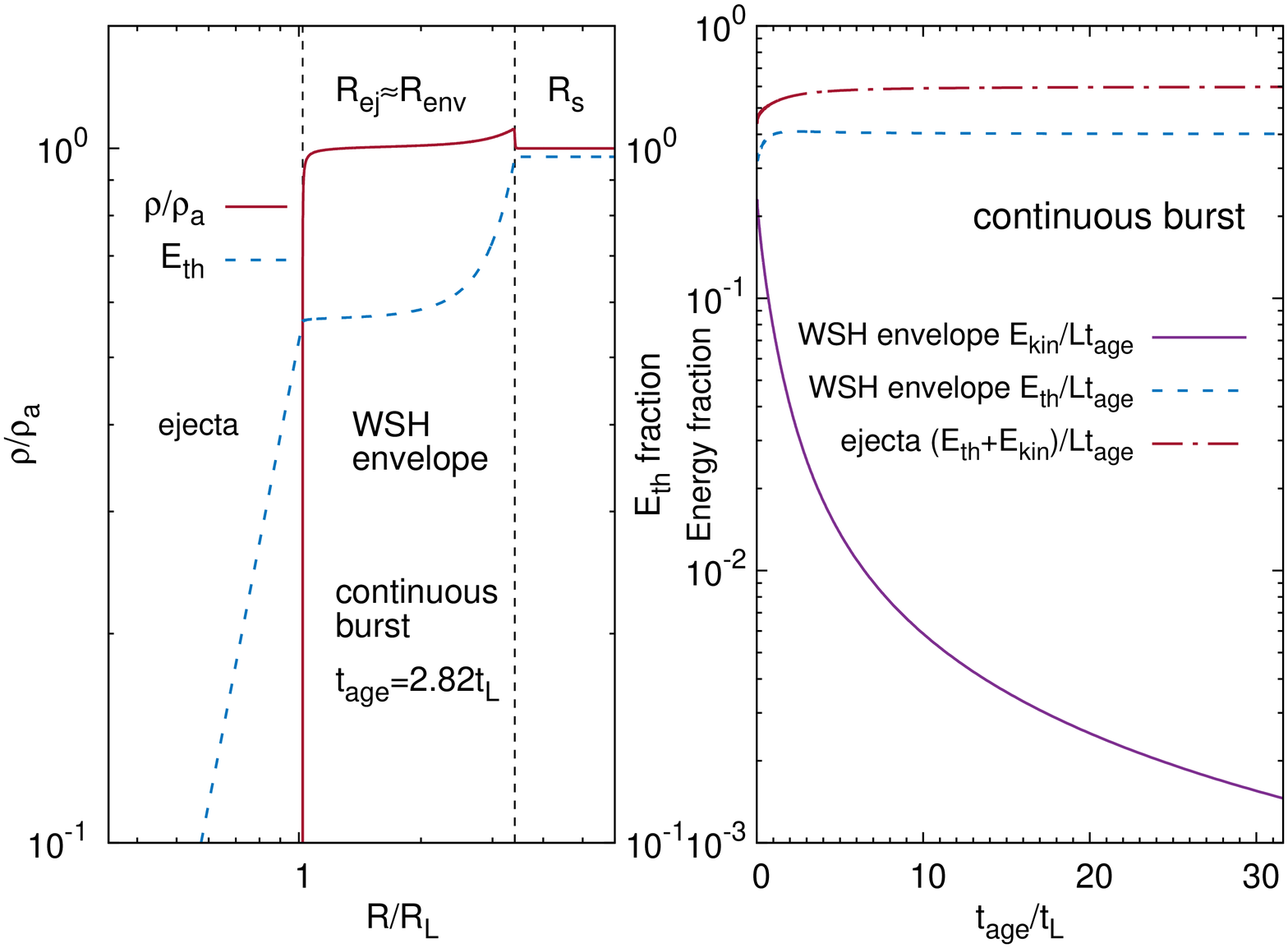} 
  \caption{ Left panel: Spatial density $\rho$ and cumulative thermal energy $E_{th}$ distribution for a continuous burst at $t_{age}=2.82t_L$. SSH envelope becomes negligible now and close to $60\%$ of injected energy $E_{inj}$ is stored in the ejecta. No WLS appears in the density and cumulative thermal energy distribution. $E_{th}$ now increases monotonically with radius.  Right panel: Time evolution of different energy components in a continuous burst with constant energy injection rate $L$. All energy components approach the values in a slow piston driven outburst as $t_{age}/t_L$ increases.} 
    \label{piston_region}
 \end{center}
 \end{figure*}

For a continuous outburst, after shock detachment its spatial structure mainly consists two regions, ejecta and WSH gas envelope, as shown in the right panel of Fig. \ref{schematic}. The SSH gas envelope now becomes negligible since the energy dissipation in a continuous outburst is negligible (only happening when $t_{age} \lesssim t_L$). In the left panel of Fig. \ref{piston_region}, we plot the density $\rho$ and accumulated thermal energy $E_{th}$ distribution for a continuous outburst with $t_{age}=2.82t_L$. The distribution of $E_{kin}$ is not shown in this figure, as for a continuous outburst the kinetic energy is negligible after the shock detachment.
The ejecta bubble in a continuous outburst has  uniform density and temperature profiles which are different from those in an instantaneous outburst.  The shock front driven by the outburst still asymptotically approaches the sound speed like in an instantaneous outburst. No WLS appears in the density and accumulated thermal energy distribution during the evolution as shown in Fig. \ref{piston_region}. Instead $E_{th}$ increases monotonically with radius in a continuous outburst which is also different from an instantaneous outburst. It is mainly because the energy injection in a continuous outburst hasn't quenched before. 

In the right panel of Fig. \ref{piston_region}, we present the time evolution of different energy components in the two regions. The shock detachment process in a continuous outburst happens at $t_{age}\sim 1t_L$ according to the energy distribution of various components. As $t_{age}/t_L$ increases, the amount of thermal energy stored in the ejecta quickly approaches the slow piston value, i.e. $E_{ej}/E_{inj}(t_{age}/t_E \rightarrow \infty)=1/\gamma_{ej}=60\%$ where $\gamma_{ej}=5/3$, while the kinetic energy carried away by the WSH gas envelope quickly shrinks to almost zero. It is clear that all the energy components presented in the figure asymptotically approach the value in a slow piston driven outburst.  Shock evolution in a continuous outburst is described in Appendix \ref{App:shock_evolution} with detail. A simple analytical approximation for shock evolution in a continuous outburst which is consistent with numerical simulation within a few percent accuracy is also provided in Appendix \ref{App:shock_evolution}.

 \subsection{Intermediate regime}
 
In the previous two subsections, we discussed the spatial structure
and energy distribution of an instantaneous outburst
($t_b/t_E\rightarrow 0$) and a continuous outburst ($t_b/t_E\rightarrow\infty$). Now, we
consider an intermediate case and study how the solution changes from
a nearly instantaneous outburst to a nearly continuous outburst as $t_b/t_E$
increases. The spatial structure and energy partition of an outburst in a uniform medium are mainly determined by two dimensionless ratios $t_b/t_E$ and $t_{age}/t_E$, where $t_E$ works like a scaling factor. As $t_{age}/t_E$ increases, the time evolution of an outburst is characterized by the transition from a strong shock to a weak shock and eventually a sound wave (see discussion in section \ref{sec:simple} and Appendix \ref{ap:b0}). 

In this section, we focus on the energy partition during such transition, especially the asymptotic behavior, for different $t_b/t_E$. For
numerical purposes we take $t_{age}=10 t_E$ in lieu of
the definition $t_{age}\rightarrow \infty$ and vary $t_b/t_E$ from 0 to 10. At
$t_{age}=10 t_E$ the shock is already very weak and further dissipation
at the shock front should not change the energy partition significantly. The spatial structure of an outburst with different $E_{inj}$ (related to $t_E$) and $t_{b}$ at a fixed shock radius $R_s$ is investigated in \cite{Forman16} with the density and temperature profile of M87. Despite the fact that medium in M87 is non-uniform, their results are consistent with our discussions here if proper $t_E$ and $R_E$ are chosen. Time evolution and energy partition of an outburst in a non-uniform medium will be discussed in detail in section \ref{non-uniform}.

\begin{figure}
 \begin{center}
 \includegraphics[width=\columnwidth]{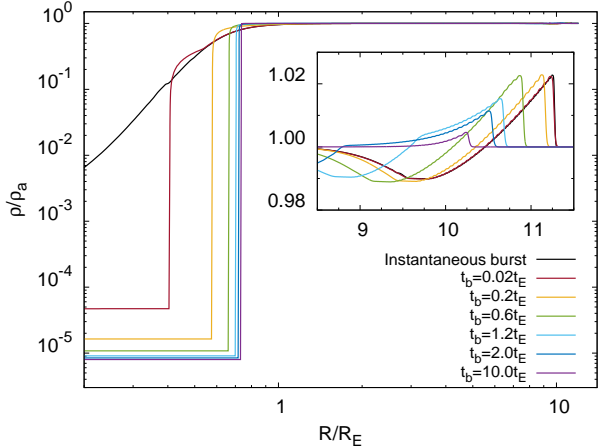} 
  \caption{Normalized radial density profiles in logarithmic scale for an outburst at
    $t_{age}=10t_E$ for different values of $t_b/t_E$, illustrating
    the transition from a ``short/fast'' outburst to a ``long/slow''
    outburst. When $t_b/t_E\lesssim 0.5$, the structure of inner part
    has a strong dependence on the value of $t_b/t_E$ while the outer
    part remains almost the same. The inlay shows the same profiles in linear scale,
    but focusing on the outer part near the shock front. In
    contrast to the inner structure, the outer part is most sensitive to the
    values of $t_b/t_E\gtrsim 0.5$.}
    \label{various_td_density}
 \end{center}
 \end{figure}

A change in the $t_b/t_E$ parameter causes strong rearrangements in the
inner and outer parts of the affected volume. Here the inner part refers to the ejecta and SSH envelope while the outer part corresponds to WSH envelope including the WLS. We illustrate this in
Fig.~\ref{various_td_density}, which shows the logarithmic density
profile of the inner part, while the inlay presents the density profile
of the outer part in linear scale. Based on these plots we can
introduce a characteristic value of $t_b/t_E\sim 0.5$ which separates
the regime of  ``short/fast'' outbursts from that of ``long/slow'' outbursts.

In the short/fast outburst regime ($t_b\lesssim 0.5t_E$), the outer part is weakly sensitive to
$t_b/t_E$ and resembles the outcome of an instantaneous outburst. In contrast, the structure of the inner part is very sensitive to the
$t_b/t_E$ ratio. As shown in
Fig.~\ref{various_td_density}, the size of the ejecta bubble
and the energy stored in it increases very quickly with $t_b/t_E$. At the same time, the amount of
gas in the SSH envelope and the energy stored within it decreases quickly with $t_b/t_E$, and
becomes negligible for $t_b/t_E\gtrsim 1$. Thus, the key signature of a
short/fast outburst is the presence of a SSH envelope, i.e. hot and low
density shell around the ejecta. We assume that observationally the
ejecta can be identified as a boundary of the radio-bright region. The
presence of the SSH envelope just outside ejecta could reveal itself
in X-rays as a low surface brightness ring. To the best of our
knowledge such structures have not been found in real objects.

In the slow/long outburst regime ($t_b/t_E \gtrsim 0.5$), the inner and outer parts respond to the change of $t_b/t_E$ in a completely opposite way. While the inner region is
insensitive to the value of $t_b/t_E$, the outer region changes
considerably as shown in Fig. \ref{various_td_density}.  As $t_b/t_E$
increases, the Mach number of the shock front decreases and the amount
of energy carried away by the WLS drops quickly with increasing
$t_b/t_E$ and eventually becomes negligible.

\subsection{Energy partition in AGN driven outburst}{\label{sec:energypartition}}
For an AGN driven outburst in galaxy or cluster, we are particularly
interested in the following question - what fraction of the injected
outburst energy $E_{inj}$ could escape from the system (i.e., escape
to very large distances), and what fraction could contribute to
balancing the gas cooling loss in the cluster core? After the shock
detachment, the ejecta and the shock heated gas envelope gradually slow down and
their contribution to the kinetic energy becomes small. All the kinetic energy
is instead associated with the outgoing (weak) shock and the gas
behind it. As $t_{age}/t_E \rightarrow \infty$, the shock transforms into a
sound wave and the value of $E_{kin}$ approaches its asymptotic
value. The energy carried away by the wave, which eventually escape from
the system, would then simply be $E_{out}=2E_{kin}$. The remaining part
$E_{inj}-E_{out}$ is distributed in the ejecta and shocked materials. When $t_{age}\gg t_E$, all motions are already
subsonic and no further dissipation is taking place. The amount of energy that ejecta and the swept up gas can release as radiation losses, when returning to its initial state, can be estimated as follows. Consider a gas shell with initial pressure $P_0=P_a$ and volume $V_0$. During the outburst, the pressure and volume of the shell becomes $P_c$ and $V_c$ respectively. Now imagine that at first the gas returns to its
original pressure $P_0$ quickly enough, so that radiative cooling can be neglected. At this moment, the volume of the gas is $V_1=V_c(P_c/P_0)^{1/\gamma}$, where $\gamma$ is the specific heat index of the gas shell. The gas then radiatively cools and recovers the initial
volume $V_0$ while maintaining the pressure equilibrium with the environment. The amount of energy released through radiative cooling
is then
\begin{equation}
  H=\frac{\gamma}{\gamma-1}P_0V_0\left [\left (\frac{P_c}{P_0}\right)^{1/\gamma}  \left( \frac{V_c}{ V_0}\right)-1\right ].
  \label{eq:h}
\end{equation}
We can now calculate the contributions to this energy from the swept up material $H_{swept}$ and ejecta $H_{ej}$ by integrating eq.~(\ref{eq:h}) over their volumes at any moment of the simulations when the phase of shock heating is over.  

For the swept up material, the value of $H_{swept}$ characterizes the
amount of ICM heating that has already happened during the shock
propagation through the gas. In the case of a long or continuous
outburst ($t_b\gg t_E$), the amount of
shock heating is small comparing with $E_{inj}$ at time $t_{age}\gg t_E$ and therefore $H_{swept}\ll E_{inj}$. In
the opposite limit of an instantaneous outburst ($t_b\ll t_E$), $H_{swept}$ is
expected to be comparable to $E_{inj}$.

For ejecta, the initial volume $V_0$ is negligible. Therefore
\begin{equation}
  H_{ej}\approx\frac{\gamma_{ej}}{\gamma_{ej}-1}P_0V_c\left (\frac{P_c}{P_0}\right)^{1/\gamma_{ej}}.
  \label{eq:hej}
\end{equation}
Furthermore, at $t_{age}\gg t_E$, the pressure in the
entire central region relax to a value very close to the initial pressure $P_0$,
i.e., $P_c\approx P_0$. As a result, the above equation simply reduces to
$H_{ej}\approx \frac{\gamma_{ej}}{\gamma_{ej}-1}P_0V_c$, i.e. enthalpy
of the ejecta. Although this energy is associated with the ejecta and
does not correspond to the ICM heating, it has been argued
that if buoyant ejecta rise in the cluster atmosphere over several
pressure scale-heights, then its enthalpy could be transferred to the
ICM and used for ICM heating \citep[e.g.,][]{2001ApJ...554..261C,2002MNRAS.332..729C,2001ASPC..250..443B}.
Unlike $H_{swept}$, the value of $H_{ej}$ is
expected to be large for a long outburst and small for a
short outburst.

\begin{figure}
 \begin{center}
 \includegraphics[width=\columnwidth]{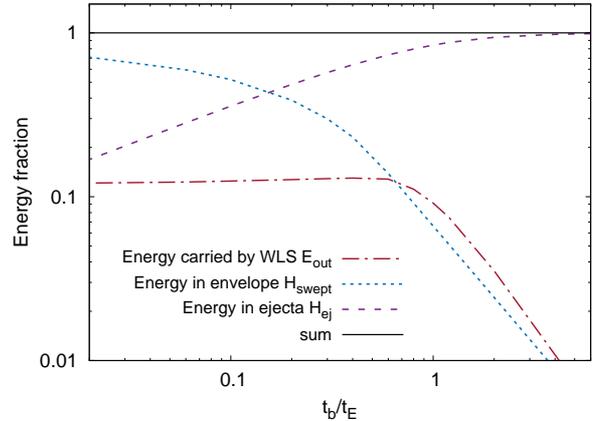} 
  \caption{Dependence of energy partition among various components on
    the ratio $t_b/t_E$. The partition of energy is evaluated at
    $t_{age}/t_E=10$ which could be considered as a good approximation for the asymptotic limit
    at $t_{age}/t_E\rightarrow \infty$. The small values of $t_b/t_E$
    correspond to a short outburst, with significant
    fractions of energy in all components. As $t_b/t_E$ increases the
    limit of slow piston driven outburst is attained. }
    \label{various_td}
 \end{center}
 \end{figure}

The above arguments are illustrated in Fig. \ref{various_td}, where we
plot $E_{out}, H_{swept}$ and $H_{ej}$ evaluated at $t_{age}=10t_E$ for
different $t_b/t_E$. All values are normalized by the injected energy
$E_{inj}$. The sum $(E_{out}+H_{swept}+H_{ej})/E_{inj}$ equals to
unity and satisfies the condition of energy conservation, which
indicates that our calculations are self-consistent. As shown in
Fig. \ref{various_td}, when $t_b/t_E<0.5$, the ratio
$E_{out}/E_{inj}\approx 0.12$ remains constant and is consistent with
the value for an instantaneous outburst. When $t_b/t_E>0.5$, the
amount of escaping energy drops very quickly as $t_b/t_E$
increases. When $t_b/t_E>4$, the amount of energy that could escape
from the system already drops below $1\%$ of $E_{inj}$ and the
evolution of the outburst becomes close to a slow piston driven/continuous outburst. The contribution of $H_{swept}$ to the total
energy is expected to be maximal for an instantaneous outburst, in which
the contribution of $H_{ej}$ is negligible. The asymptotic value
$H_{swept}\sim 0.88 E_{inj}$ corresponds to a very short outburst
(beyond the minimal value of $t_b/t_E$ shown in the plot). As
$t_b/t_E$ increases the value of $H_{swept}$ drops and becomes less
than $10\%$ at $t_b/t_E\sim 1$. On the contrary, $H_{ej}$ steadily
increasing with $t_b/t_E$; it exceeds $50\%$ at $t_b/t_E\sim 0.2$ and
approaches unity for very long outbursts.  For the continuous outburst
model in stratified atmospheres of clusters, buoyancy forces break the
bubble and displace it when the expansion of the ejecta becomes
substantially subsonic \citep[e.g.,][]{2000A&A...356..788C}. This
corresponds to large values of $t_b/t_E$. It is therefore likely that
in clusters the main repository of released energy is the enthalpy
of the ejecta $H_{ej}$.

According to eq.~(\ref{eq:hej}), it is clear that the amount of energy stored
in the ejecta is determined by the size (volume) of the ejecta. In X-ray
observation, the ejecta can be identified as a radio-bright and
X-ray-dim region. The dimensionless ratio of the shock radius $R_s$
and the ejecta radius $R_{ej}$ then becomes an important physical quantity
characterizing the properties of the outburst. This quantity is also
easy to evaluate from observations, apart from the uncertainty caused
by the deviations from spherical symmetry. For a continuous outburst, an
analytical approximation for the $R_s/R_{ej}$ ratio is available, which is
presented in Appendix \ref{ap:d}.

\section{Non-uniform medium}\label{non-uniform}    

In this section, we use the Perseus cluster and M87 as examples to show that the dynamical evolution of an
outburst in a monotonous, smooth and shallow (power index $\lesssim2$)
density and pressure profiles is qualitatively similar to that in a
uniform medium. Moreover, we show below that for the same characteristic length $R_E$ and time $t_E$, the quantitative predictions of an outburst in uniform medium reproduce major variables with an accuracy of 20-30 per cent. It enables us to develop simple tools,
based on the simulations in a uniform medium, to estimate intrinsic
properties of an outburst in a non-uniform medium without running
numerical simulation tailored to each specific case.
  
In a uniform medium, the characteristic length and time scales are defined by eqs.~(\ref{eq:charac_R}) and (\ref{eq:charac_t}). For a more realistic case, when $P_a(r)$ and $\rho_a(r)$ vary with radius, it is natural to replace eq.~(\ref{eq:charac_R}) that defines $R_E$ with 
\begin{eqnarray}
E_{inj}=\int_0^{R_E} P_a(r) 4\pi r^2  dr .
\end{eqnarray}
For the time scale $t_E$, one could replace the sound speed $c_s$ in eq (\ref{eq:charac_t}) with volume averaged value, i.e. adopt the volume averaged temperature $\overline{T}$ to calculate $c_s$. $R_E$ and $t_E$ defined with the above method could be used to characterize the evolution of an outburst in a non-uniform medium. But for the galaxies and clusters of interest here with shallow density and pressure profiles, a simpler definition of $R_E$ and $t_E$ with pressure and density at the shock front, i.e. 
\begin{equation}
E_{inj}=\frac{4\pi}{3}P_a(R_s) R_E^3, 
\label{eq: RE_simple}
\end{equation}
and
\begin{equation}
t_E=\frac{R_E}{\sqrt{\gamma_a P_a(R_s)/\rho_a(R_s)}}, 
\label{eq: tE_simple}
\end{equation}
are found to provide results close to those derived with volume averaged quantities. In the rest of discussion, unless specifically noted we will use the definitions in eq (\ref{eq: RE_simple}) and (\ref{eq: tE_simple}) for simplification.

The primary goal of this section is to develop simple tools to map major observables to intrinsic
properties of an outburst, which could be easily applied to different objects. Before we present the tools developed according to simulation in a uniform medium, we run simulations for outbursts in
realistic cluster profiles of $P_a(r)$ and $\rho_a(r)$ taken from observations and verify that the predictions of uniform medium runs
agree with more accurate calculations.

\subsection{Numerical Simulations for M87 and the Perseus cluster}
The observed values of shock radius $R_s$, ejecta radius $R_{ej}$ and Mach number $M$ for these two objects are
listed in Table \ref{observation}. The volume averaged pressure
$\overline{P}$ and density $\overline{\rho}$ within the outburst in Table \ref{observation}
are calculated with the ``unperturbed'' profiles presented in Appendix \ref{App: M87_Perseus}.

For M87 and the Perseus cluster, we run a set of numerical simulations to
search for a set of parameters $(E_{inj},t_b,t_{age})$ which can
reproduce the observed values $(R_s,R_{ej},M)$. Plausible sets of
parameters $(E_{inj},t_b,t_{age})$ derived from these simulations are
summarized in Table \ref{simulation_nonuniform} (see Appendix \ref{App: M87_Perseus} for
details). These values are consistent with previous works
\citep{Forman07,Forman16,Zhuravleva16}.

Based on these simulations, we concluded that the physical properties of an outburst in a non-uniform medium is qualitatively
similar to that in a uniform medium (See Appendix \ref{App: M87_Perseus} for
details). The spatial structure of the outburst still contains three regions: ejecta, SSH and WSH as shown in Fig \ref{schematic}. The quantitative difference is mainly caused
by two reasons. At first, $t_E$ and $R_E$ are evolving with time $t_{age}$ for an outburst in a non-uniform medium. In a realistic galaxy or cluster profile like M87 and the Perseus cluster, $t_E$ is increasing with time. As a result, $t_b/t_E$ is decreasing with $t_{age}$ and the outburst will appear to be a shorter/faster outburst as $t_{age}$ increases. Secondly, gravity plays a role in the energy partition of an outburst. The shocked materials pushed by the central ejecta end up with a higher gravitational potential during the outburst. Therefore the amount of energy that could contribute to radiative cooling is not only determined by the enthalpy of the ejecta and shocked materials but also related to the gravitational energy of the shocked materials (gravitational energy of ejecta is negligible due to negligible mass). The energy carried away by the WLS however is not affected significantly by the non uniform environment with shallow profiles.

\begin{table}
\centering
\caption{Basic parameters of M87 and Perseus cluster from observation}
\begin{threeparttable}
\begin{tabular}{ccc}
\hline\hline
object& M87 &Perseus cluster\\
\hline
$R_s(\rm kpc)$&13.0&15.0\\
$R_{ej}(\rm kpc)$&3.0&7.5\\
$M^a$&1.2&1.16\\
$\overline{n}^b(\rm cm^{-3})$&$4.5\times 10^{-2}$&$0.1$\\
$\overline{P}^b(\rm keVcm^{-3})$&$7.8\times 10^{-2}$&$0.35$\\
$n_s^c(\rm cm^{-3})$&$3.1\times 10^{-2}$&$0.1$\\
$P_s^c(\rm keVcm^{-3})$&$5.6\times 10^{-2}$&$0.33$\\
$n_{ej}^d(\rm cm^{-3})$&$0.13$ & $0.11$\\
$P_{ej}^d(\rm keVcm^{-3})$ & $0.2$ &$0.36$\\
Reference&1,3,4,6&2,5\\
\hline\hline
\end{tabular} 
\begin{tablenotes}
\small
\item[a] Assuming $\gamma=5/3$.
\item[b] Volume averaged quantities. 
\item[c] Quantities at the shock front $R_s$.
\item[d] Quantities at the ejecta front $R_{ej}$. 
\item  1. \cite{Forman07}; 2. \cite{Graham08}; 3. \cite{Xiang09}; 4. \cite{Simionescu09}; 5. \cite{Zhuravleva16};
 6. \cite{Forman16}
\end{tablenotes}
\end{threeparttable}
\label{observation}
\end{table}

\subsection{Simple tools based on simulation in uniform medium}{\label{sec:mapping}}

In this subsection, we discuss simple recipes to estimate the
intrinsic properties $(E_{inj},t_b,t_{age})$ of an outburst based on
observables $(R_s,R_{ej},M)$ in addition to ambient density $\rho(r)$ and
pressure $P(r)$ profiles without running numerical simulation.

The age of an outburst $t_{age}$ satisfies \begin{equation}
t_{age}=\int_0^{R_s} \frac{dr}{M(r) c_s(r)}
\end{equation}
where $c_s(r)$ is the sound speed at radius $r$ and $M(r)$ is the Mach
number of the outburst when its shock front arrives
at the radius $r$. For outbursts in clusters or galaxies with shallow density and pressure profiles, the Mach number $M(r)$ is decreasing with radius $r$ and time $t_{age}$ during the evolution. Thus
the dominant contribution to the above integral comes from the stage with $r\sim R_s$ and $M\sim
M(R_s)$. Therefore we further assume 
\begin{equation}
t_{age}=\frac{\chi R_s}{M(R_s)c_s(R_s)}
\label{tage_simple}
\end{equation}
where $\chi\sim O(1)$ is a coefficient that depends on the evolution history of an
outburst. As $t_{age}/t_E \rightarrow \infty$, the shock front becomes weak with $M\rightarrow 1$ and in the meantime $\chi$ asymptotically approaches the value of unity. According to above discussion, for outbursts with weak shock a simple estimate for $t_{age}$ would be $t_{age}=R_s/M(R_s)c_s(R_s)$ as $\chi$ is expected to be close to 1 in such a situation. 

\begin{figure}
 \begin{center}
 \includegraphics[width=\columnwidth]{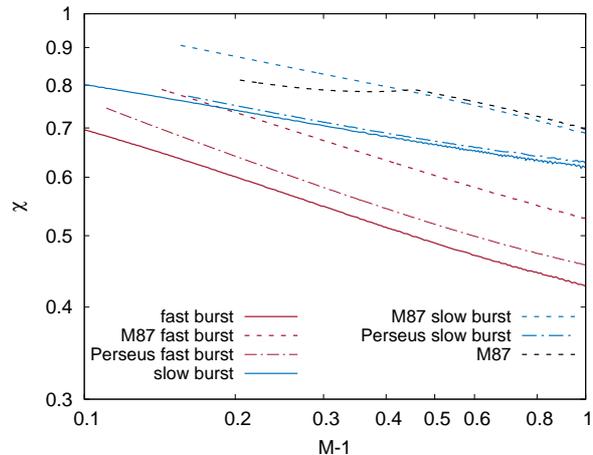} 
  \caption{ $\chi$ as a function of $M-1$. $\chi$ is defined in eq (\ref{tage_simple}) and could be considered as the ratio between simulated $t_{age}$ and the estimation from simple analytical formula $t_{age}=R_s/M(R_s)c_s(R_s)$. $M(R_s)$ and $c_s(R_s)$ are the Mach number and sound speed at shock radius $R_s$. Solid, dashed and dotted-dashed lines are for simulation results in uniform medium, realistic profiles in M87 and realistic profiles in Perseus cluster respectively. Red lines are for fast instantaneous outburst while blue lines are for slow continuous outburst. For fast burst in M87 and Perseus cluster, we assume $E_{inj}=5.5\times 10^{57}erg$ and $1.1\times 10^{59}erg$ respectively according to our best fit model. Black line is an intermediate outburst which represents our best fit model for M87. Blue dotted dashed line shows our best fit model for Perseus cluster. In the calculation of blue dashed line we apply $L\approx 7\times 10^{43}\ergs$ which is comparable with $L$ from our best fit model for M87. }
    \label{age_estimate}
 \end{center}
 \end{figure}

In Fig. \ref{age_estimate}, we plot $\chi$ as a function of $M-1$ for outbursts with different $t_b/t_E$ in both uniform media and realistic galaxy profiles. It is found that, for a given Mach number, $\chi$ in a slower outburst is more close to 1. It is not a surprise, as shock in a slow outburst is weaker at early time. Besides $\chi$ derived from a non-uniform medium is larger than that from a uniform medium. It is mainly because Mach number in a realistic galaxy or cluster profile drops more slowly with time due to the density gradient. As a result, the outbursts spend more time in the small $M$ regime, making $\chi$  closer to 1. For slow outbursts with weak shocks $M  \lesssim 1.5$, it is found that $\chi$ is close to 1 with only about $20\%-30\%$ offset as shown in Fig. \ref{age_estimate}. So we could simply apply eq (\ref{tage_simple}), assuming $\chi=1$, to estimate $t_{age}$. More accurate $t_{age}$ can be obtained if we use for example the continuous outburst line (blue solid) to correct $\chi$ for a slow outburst ($t_b/t_E \geq 0.5$) and the instantaneous outburst line (red solid) to correct $\chi$ for a fast outburst ($t_b/t_E< 0.5$). For M87 and the Perseus cluster, $t_{age}$ estimated with eq (\ref{tage_simple}) are presented in Table \ref{simulation_nonuniform} and are consistent with numerical simulation results with $20\%-30\%$ accuracy.

\begin{figure*}
 \begin{center}
 \includegraphics[width=0.9\textwidth]{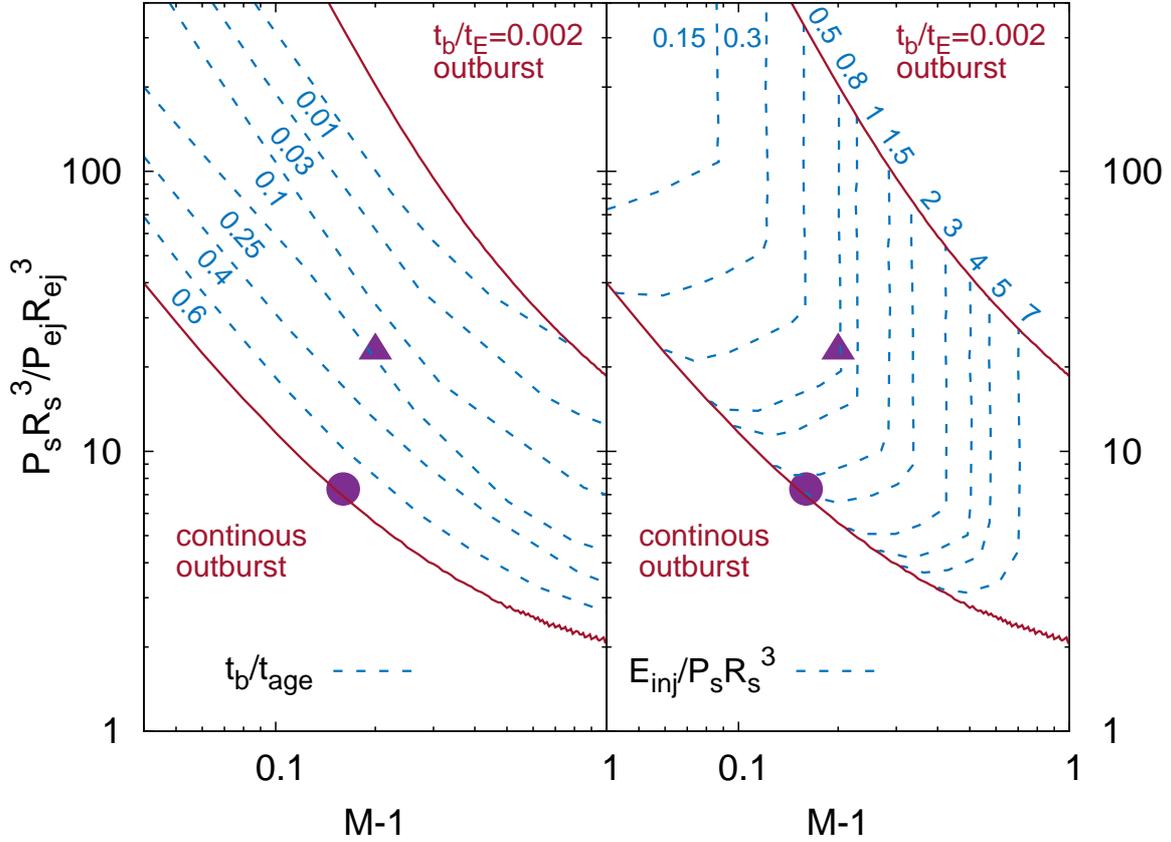} 
  \caption{x axis: $M-1$ where $M$ is the Mach number of the shock wave; y axis: $P_sR^3_s/P_{ej}R_{ej}^3$ where $R_s$ is shock radius, $R_{ej}$ is ejecta radius, $P_{ej}$ is the pressure at $R_{ej}$ and $P_s$ is the pressure at $R_s$. Left panel: Red solid lines from top to bottom represent a fast outburst with $t_b/t_E=0.002$ and a continuous outburst respectively. Blue dashed lines is the contour with different $t_b/t_{tage}$ ratios. Note the red continuous outburst line also indicates the line with $t_b/t_{tage}=1$. Right panel: Red solid lines from top to bottom again represent a fast outburst with $t_b/t_E=0.002$ and a continuous outburst, respectively. Blue dashed lines show the contour with different $E_{inj}/P_sR^3_s$ ratios where $E_{inj}$ is the total injected outburst energy. Purple triangle and circle are for M87 and Perseus cluster respectively based on numerical simulation results in Table \ref{simulation_nonuniform}. The size of the symbol is not a indication for the error bar.} 
    \label{PVM}
 \end{center}
 \end{figure*}

Next let us discuss a possible way to derive $t_b$ and $E_{inj}$ based on available
observables including shock radius $R_s$, ejecta radius $R_{ej}$, Mach number $M$ and the pressure profile $P(r)$. To facilitate the
comparison between observables and the intrinsic properties of an outburst, we apply two dimensionless numbers $M$ and $G$ to characterize the state of the system at a given
time $t_{age}$. The first parameter $M$ is simply the Mach number of the
shock and the second parameter
\begin{eqnarray}
G=\frac{P_sR_s^3}{P_{ej}R_{ej}^3},
\end{eqnarray}
is the ratio of thermal energies within the volumes defined by the
shock and ejecta radii, $R_s$ and $R_{ej}$, respectively. Note in the above equation we replace the integration of the energy density over volume with the product of the
volume and the pressure at the corresponding radius
to further simplify the comparison. We
then run a group of simulations in a uniform medium with different
values of $t_b/t_E$. According to the simulation results, we further build map between $(M,G)$ and dimensionless ratios $t_b/t_{age}$ and $E_{inj}/P_sR_s^3$, which could be used to calculate $t_b$ and $E_{inj}$ easily. In Fig. \ref{PVM}, we present the contours of $t_b/t_{age}(M,G)$ (left panel) and $E_{inj}/P_sR_s^3(M,G)$ (right panel) as a function of $M-1$ and $G$.
The bottom red solid line
corresponds to a continuous outburst, while the top red solid line
corresponds to a short outburst with $t_b/t_E=2\times 10^{-3}$. During the evolution of an outburst, the Mach number $M$ decreases with $t_{age}$, so in Figs.~\ref{PVM} time
increases from right to left. As long as
$t_{age}\leq t_b$ the time evolution of an outburst follows the continuous outburst
line. Once the outburst quenches, it gradually switches to a regime in which the shock radius linearly
increases with time and the ejecta radius remains almost static.

Next we apply the map shown in Fig.~\ref{PVM} to estimate $t_b$ and $E_{inj}$ for M87 and the Perseus cluster based on observables presented in Table \ref{observation}. The results are shown in Table \ref{simulation_nonuniform} under the name of the mapping method.
$E_{inj}, t_b$ and $t_{age}$ derived from our simple mapping method are found to be
consistent with numerical simulations tailored for
M87 and the Perseus cluster with a $20\%-30\%$ accuracy. The estimate for the Perseus cluster is better than M87, because the outburst in Perseus cluster is still in the core region with flat density and pressure profiles. As long as the density and pressure distribution in an outburst are not very steep (with power law index
$\lesssim 2$), the mapping method discussed above according to simulation in a uniform medium should be able to provide a good estimate for $E_{inj}, t_b$ and $t_{age}$. The validity of the map shown in Fig. \ref{PVM} is illustrated in Appendix \ref{App:PVM_prove} in detail. 

Another way to estimate $t_b$ and $E_{inj}$ based on known $t_{age}$ is to use Fig. \ref{various_td} which can be used as an independent check for the mapping method. 
It requires a bit more calculation. In order to derive $E_{inj}$, we further introduce two physical variables,
\begin{equation}
E_{c}= \frac{4\pi R_{ej}^3 P(R_{ej})}{3(\gamma_{ej} -1)}
\end{equation}
and
\begin{equation}
E_{a}=(M-1)^2 \frac{4\pi (R_s^3-R_{ej}^3) P(R_s)}{3(\gamma_a -1)},
\end{equation}
where $R_{ej}$ and $R_s$ are the ejecta and shock radius respectively.
$E_c$ is a approximation for the thermal energy stored in the ejecta bubble while $E_a$ characterizes the thermal energy transferred into the shocked ambient medium.

For a continuous outburst, the injected energy $E_{inj}\approx\gamma_{ej} E_{ej}\approx\gamma_{ej} E_c$ according to the discussion in section \ref{uniform medium}. 
For a very short outburst, the bulk of the injected energy $E_{inj}$ is transferred to the shocked ambient medium. We instead find $2\gamma_a E_{a}$ provides a good approximation of $E_{inj}$ for the age range $(0.1t_E,2t_E)$ we're interested in here. By combining the expressions for $E_{inj}$ from the above two limiting cases, we end up with the following simple formula 
\begin{equation}
E_{inj}=\left\{ \begin{array}{ll}
\gamma_{ej}E_c+\gamma_a E_a, \,\quad \mbox{if } E_c\geq E_a\\
2\gamma_a E_a, \qquad\qquad \mbox{if } E_c< E_a\\
\end{array}\right.
\end{equation}
to estimate $E_{inj}$.
Once we obtain $E_{inj}$, we can further derive $t_b/t_E$ by comparing $\gamma_{ej} E_{c}/E_{inj}$ with the purple dashed line in Fig. \ref{various_td} which represent the percentages of $\gamma_{ej} E_{ej}$. By using eq (\ref{eq: tE_simple}) to derive $t_E$, we can then calculate $t_b$ easily. $t_b$ and $E_{inj}$ derived with this method are presented in Table \ref{simulation_nonuniform} under the name of the analytical estimate. The results from analytical estimates are consistent with the mapping method and numerical simulations for both M87 and the Perseus cluster. In the calculation of $t_b$ for the Perseus cluster, we found $t_b>t_{age}$ which is not physical. It is mainly because the results shown in Fig. \ref{various_td} are intended for asymptotic values of the energy partition. The observed outbursts however may not have relaxed to such asymptotic values yet. As a result, we obtain an unreasonably large $t_b$. In such a situation, we simply assume $t_b=t_{age}$.  

Note both the mapping method and the analytical estimate described in this section are designed for the late time evolution of an outburst when the shock front already detaches from the ejecta bubble and becomes a weak shock. Both the Perseus cluster and M87 are found to satisfy this condition. It is probably a selection bias, as outbursts at earlier times shall have smaller sizes which would be harder to resolve in observations. The dynamics of an outburst in early time will have  stronger dependence on the asymmetry of the central AGN which is probably beyond the simple model discussed here.

 \begin{table*}
\centering
\caption{Basic parameters of M87 and Perseus cluster from different methods by fitting the outburst parameters in Table \ref{observation}.}
\resizebox{\textwidth}{!}{
\bgroup
\def\arraystretch{1.5}%
\begin{threeparttable}
\begin{tabular}{lc|ccc|ccc}
\hline\hline
Object&Method&$E_{inj}(\rm 10^{58}erg)$&$t_b(\rm Myr)$&$t_{age}(\rm Myr)$&$E_{inj}/P_sR_s^3$&$t_b/t_E$&$t_b/t_{age}$\\
\hline\hline
&Numerical simulation&$0.55$&2.0&12.3&0.95&0.18&0.16\\
M87&Mapping method&$0.46$&$3.3(2.4)$&15(11)&$\sim 0.8$&0.29(0.21)&$\sim 0.22$ \\
&Analytical estimate&0.5&$2.8$&15(11) &0.86&$\sim 0.25$&$0.18(0.24)$ \\
\hline
&Numerical simulation&10.8&10.3&10.3&2.1&0.83&1 \\
Perseus &Mapping method& $9.3$&11.7(9)&13(10)&$\sim 1.8$&0.94(0.73)&$\sim 0.9$\\
cluster&Analytical estimate& 8.9 &$11.8$&13(10)&1.7&$\sim 0.95$&$0.9$(1)\\
\hline
\end{tabular}

\begin{tablenotes}
\small
\item For simplicity, $t_E$ is calculated with pressure $P$ and density $\rho$ at the shock front $R_s$. 
\item Quantities in brackets are estimated with $\chi=0.75$ based on the continuous outburst line in Fig.~\ref{age_estimate}, while quantities outside the brackets are calculated with $\chi=1$. 
\item Note for analytical method there is conflict between $t_b$ and $t_{age}$ in Perseus cluster, see text for detail. 
\item "$\sim$" symbol indicates quantities read from figures.
\end{tablenotes}
\end{threeparttable}
\egroup
}
\label{simulation_nonuniform}
\end{table*}


\section{Classification of outbursts by duration and age}
\begin{figure}
 \begin{center}
 \includegraphics[width=\columnwidth]{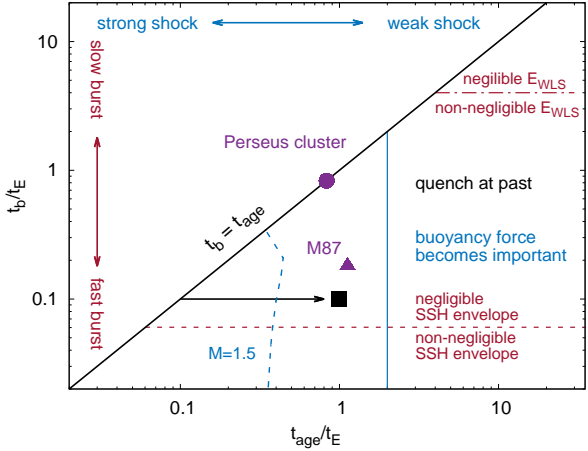} 
  \caption{Schematic figure for outbursts in galaxy or cluster. Horizontal axis is the $t_{age}/t_E$ ratio which indicates the time evolution. Vertical axis is the $t_b/t_E$ ratio which represents the transition from fast outburst to short outburst. The black solid line from bottom left to top right shows the evolution track of an continuous outburst, i.e $t_{age}=t_b$. The dashed blue (almost vertical) line shows the position with Mach number $M=1.5$, which characterizes the transition from strong shock to weak shock (left to right). The blue solid vertical line schematically
shows the region where in real clusters buoyancy effects become
important (to the right from this line). The red dashed line separates the region with negligible SSH envelope and the region with non-negligible SSH envelope. Above the red dotted-dashed line, the energy carried out by the WLS structure already drops below $1\%$ of $E_{inj}$ and becomes negligible. Purple circle and triangle show the position for Perseus cluster and M87 respectively according to simulation results. Black square is an outburst with $t_b/t_E=0.1$ and $t_{age}/t_E=1$. The black arrow next to it indicates the time evolution of the outburst after it quenches at $t_b=0.1t_E$. }
    \label{schematic_data}
 \end{center}
 \end{figure}

{\label{sec:classification}}
In the previous sections, we have already shown that the intrinsic properties of an outburst, i.e. $t_b$, $t_E$ and $t_{age}$, can be derived from major observables, i.e. $R_s, R_{ej}$ and $M$ plus density and pressure profiles of the galaxy or cluster.  Here we demonstrate the classification of outbursts in a two-dimensional plot according to their physical properties (see Fig.~\ref{schematic_data}). The horizontal axis is the dimensionless ratio $t_{age}/t_E$ which characterizes the time evolution of an outburst. As $t_{age}/t_E$ increases, the shock front in an outburst gradually transitions from a strong shock to a weak shock.  The vertical axis is the dimensionless ratio $t_b/t_E$, which is mainly determined by the energy injection history of an outburst. As $t_b/t_E$ increases, a short/fast outburst gradually changes into a long/slow outburst. 
In Fig.~\ref{schematic_data}, $t_{age}$ is simply the time elapsed since the onset of
an outburst. The definition of $t_b$ and $t_E$ in this plot however requires
extra explanation that is given below. In observations of a realistic object, it either quenched at past with $t_b<t_{age}$ or has ongoing energy injection with $t_b=t_{age}$. So in Fig.~\ref{schematic_data}, we are only interested in the region with $t_b\leq t_{age}$. The black solid line from bottom left to top right represents the evolution track of an ongoing outburst. The characteristic time $t_E$ is determined by the energy released since the onset of the outburst till $t_{age}$. Therefore, for a constant energy injection rate $L$,
$ E_{inj} = L\times {\rm min}(t_{age},t_b)$, which implies $t_E$ increases with time $t_{age}$ before the outburst quenches. According to the above definitions, the time evolution of an outburst with $t_{b}/t_E=0.1$ and $t_{age}/t_E=1$ (black square in Fig.~\ref{schematic_data}) would first proceed along the line
$t_{age}=t_b$ until $t_{age}=t_{b}=0.1t_E$ and then follow the horizontal black arrow with
$t_b=0.1t_E$ until $t_{age}=1t_E$ to arrive at its current position in the plot. 

Next we discuss the various regions presented in Fig.~\ref{schematic_data} to demonstrate various processes happening during the outburst. The dashed blue (almost vertical) line in Fig.~\ref{schematic_data} is the Mach number $M=1.5$ line which is considered as an indicator for the transition between strong shock and weak shock. The blue solid vertical line schematically shows the region (to the right from this line) where in real clusters buoyancy effects become important. The value of $t_{age}/t_E$
corresponding to this line is only indicative, since it depends on
the interplay between the buoyancy and expansion velocity of the
bubbles \citep[e.g.,][]{2000A&A...356..788C}. The main characteristic
of this regime is that the expansion velocities of ejecta become
substantially subsonic. The two red horizontal lines define three regions along
$t_{b}/t_E$. Below the lower dashed line (very short outburst), the shock-heating is
very significant and a prominent SSH gas
envelope is formed and possibly detected in observations. Above the upper dotted-dashed line (very long outburst),
the WLS is less significant and the energy carried away by the WLS already drops to a value below about $1\%$ of $E_{inj}$. In real clusters this region is likely inaccessible since
the buoyancy forces will disrupt the bubble before the outburst can
evolve to this stage. In the region between the two red lines, shock heating is not important, but the WLS is still present.

We mark the approximate locus of M87 (triangle) and the Perseus cluster (circle) outbursts in this
plot based on our best fit numerical simulation model shown in Table
\ref{simulation_nonuniform}. Along the $t_{age}/t_E$ axis, both observed
outbursts are located in the region with $0.3\lesssim
t_{age}/t_E\lesssim 2$. It is not surprising, since selection effects
make it difficult to detect an outburst in the early state
with small $t_{age}/t_{E}$. Detection of outbursts in the very late stage ($t_{age}/t_{E}\gg 1$) is also unlikely as the buoyancy effects start to become important with time and can
remove the ejecta (bubble) from the core region. The
observed $R_s, R_{ej}$ and $M$ for the Perseus cluster suggest an ongoing outburst while the observed parameters for M87 imply its outburst has quenched long ago. These results
are in line with previous findings
\citep[e.g.,][]{Forman07,Forman16,Zhuravleva16}. Along the $t_b/t_E$ axis
both outbursts fall into the range where shock heating is not very
important and a small fraction of energy $\lesssim 12\% E_{inj}$ is carried by the sound
wave. Most of the energy released in the outburst is captured by the
enthalpy of the ejecta/bubble.

So far we have placed only 2 objects in this plot. In the future, with
enlarged samples of objects we could possibly extract information on
the duty circle of the outbursts in galaxies and clusters by studying
their distribution in the $t_b/t_E$ v.s. $t_{age}/t_E$ plot.

\section{Conclusions}
 \label{sec:conclusion}
We have considered the evolution of a spherically symmetric outburst in
a uniform medium with non-negligible pressure, having in mind AGN feedback in the intra-cluster
medium (ICM). We, in particular, focused on the spatial structure and energy partition of the outburst during the transition from an early ST phase to a late wave-like phase.

The physical properties of an outburst in a homogeneous medium with pressure
$P_a$ and density $\rho_a$ at age $t_{age}$ are mainly determined by two quantities, the total injected energy $E_{inj}$ and the duration $t_b$. $E_{inj}$, $P_a$ and $\rho_a$ together define the characteristic length scale $R_E$ and time scale $t_E$ of the system, as given by eqs. (\ref{eq:charac_R}) and (\ref{eq:charac_t}) respectively. The ratio $t_b/t_E$ characterizes the energy injection history of the outburst. As the $t_b/t_E$ ratio increases, the system changes from an instantaneous outburst with $t_b/t_E\ll 1$
to a continuous one with $t_b/t_E\gg 1$. In the former case, much of
the energy ($\sim 88\%$) goes into the shock heated gas (at $t_{age}/t_E\gg 1$), while a small fraction of energy $\lesssim 12\%$ is carried away by the sound waves.  The enthalpy of
the ejecta is negligible. In the latter case, essentially 100\% of
energy is stored as the enthalpy of very hot ejecta, while the amount
of energy spent on shock heating of the ambient medium or outgoing
sound waves is negligible. 

The enthalpy of the bubble, i.e. $H_{ej}\approx \frac{\gamma_{ej}}{\gamma_{ej}-1}P_0V_c$ \citep{2002MNRAS.332..729C}, is usually adopted by observers to estimate the injection energy of an outburst $E_{inj}$. Based on discussion in \ref{sec:energypartition}, the ratio $H_{ej}/E_{inj}$ varies from zero to one as $t_b/t_E$ changes from zero to infinity. It is found that the outbursts in M87(Virgo) and Perseus cluster have $t_b/t_E\approx 0.2$ and $0.8$ respectively. According to Fig. \ref{various_td}, if we assume $\gamma_{ej}=4/3$, $E_{inj}/P_0V_c$ varies from 5 to 13 as $t_b/t_E$ decreases from 1 to 0.1, which is consistent with the results in \cite{Birzan04}.  

For the mechanical AGN feedback in cluster cores, 
the latter case ($t_b/t_E\gg 1$) seems to be more relevant. Indeed, under  assumption of quasi-continuous energy release, the life-time of the bubbles is set by the competition of the buoyancy and the
expansion of ejecta due to energy release \citep[e.g.,][]{2000A&A...356..788C}. Since
this balance is achieved when velocities are substantially subsonic,
this corresponds to $t_b/t_E\gtrsim 2$, where the enthalpy of the
ejecta captures the dominant fraction of the energy.

We further show that for a spherically symmetric outburst the maximum
amount of energy that is eventually carried away by a sound wave does
not exceed $\sim$12\%. This value is an upper limit, since we have
neglected effects of conduction or viscosity that may further
attenuate outward-going sound waves
\citep[e.g.,][]{Graham08}.  We will study these effects in
future work. Note that, for outbursts/outflows with strong anisotropy, the fraction of energy that goes into sound waves can be larger.

We also develop two simple methods to connect the observable $(R_{ej}, R_s, M)$ to the intrinsic properties of an outburst $(E_{inj}, t_b, t_E)$ based on the simulations of an outburst in a uniform medium. We then tested the accuracy of these approaches using numerical simulations incorporating realistic radial density and temperature profiles in M87/Virgo and the Perseus cluster. It is found that energetic outbursts in galaxies or clusters with a shallow profile (power law index $\lesssim 2$) are qualitatively similar to those in a uniform medium. The quantitative difference is also not very significant. Our approach recovers the values of $(E_{inj}, t_b, t_E)$ from simulations tailored for these objects with $20\%-30\%$ accuracy. 

The main limitation of the above approach is the assumption of spherical symmetry. One can expect this approach to work reasonably well for energy-driven outflows, while for momentum-dominated jets the spherical symmetry may be seriously violated. The diffuse morphology of typical Fanaroff-Riley type I radio sources in many cluster cores suggests that the energy-driven outflow is a reasonable assumption and therefore our conclusions hold at least approximately.

\section*{Acknowledgments}
We are grateful to the anonymous referee for useful comments which help to clarify some points in the manuscript.
We would like to thank Roger Chevalier for many useful comments, which helped us to improve the manuscript.
We're also grateful to Rashid Sunyaev and William Forman for useful discussions and collaborations on related projects.


\begin{thebibliography}{99}

 \bibitem[\protect\citeauthoryear{Begelman}{2001}]{2001ASPC..250..443B} Begelman M.~C., 2001, ASPC, 250, 443 
 
\bibitem[Bethe et al.(1958)]{Bethe58} Bethe, H. A., Fuchs, K., Hirschfelder, J. O., Magee, J. L., Neumann, R. V., 1958, Blast Wave (No. LA-2000), Los Alamos National Lab NM
 
\bibitem[B{\^i}rzan et al.(2004)]{Birzan04} B{\^i}rzan, L., Rafferty, D.~A., McNamara, B.~R., Wise, M.~W., \& Nulsen, P.~E.~J.\ 2004, \apj, 607, 800 
 
 
\bibitem[Caramana et al.(1998)]{Caramana98} Caramana, E.~J., Shashkov, M.~J., \& Whalen, P.~P.\ 1998, Journal of Computational Physics, 144, 70 

\bibitem[Chugai et al.(2011)]{CCS11} Chugai, N.~N., Churazov, E.~M., \& Sunyaev, R.~A.\ 2011, \mnras, 414, 879 


\bibitem[\protect\citeauthoryear{Churazov et al.}{2000}]{2000A&A...356..788C} Churazov E., Forman W., Jones C., B{\"o}hringer H., 2000, A\&A, 356, 788 

\bibitem[\protect\citeauthoryear{Churazov et al.}{2001}]{2001ApJ...554..261C} Churazov E., Br{\"u}ggen M., Kaiser C.~R., B{\"o}hringer H., Forman W., 2001, ApJ, 554, 261 

\bibitem[\protect\citeauthoryear{Churazov et al.}{2002}]{2002MNRAS.332..729C} Churazov E., Sunyaev R., Forman W., B{\"o}hringer H., 2002, MNRAS, 332, 729 

\bibitem[\protect\citeauthoryear{Churazov et al.}{2003}]{Churazov03} Churazov, E., Forman, W., Jones, C., \& B{\"o}hringer, H.\ 2003, \apj, 590, 225 


\bibitem[Dokuchaev(2002)]{Dokuchaev02} Dokuchaev, V.~I.\ 2002, \aap, 395, 1023 

\bibitem[\protect\citeauthoryear{Fabian et al.}{2006}]{2006MNRAS.366..417F} Fabian A.~C., Sanders J.~S., Taylor G.~B., Allen S.~W., Crawford C.~S., Johnstone R.~M., Iwasawa K., 2006, MNRAS, 366, 417 

\bibitem[\protect\citeauthoryear{Fabian}{2012}]{2012ARA&A..50..455F} Fabian A.~C., 2012, ARA\&A, 50, 455 

\bibitem[Forman et al.(2007)]{Forman07} Forman, W., Jones, C., Churazov, E., et al.\ 2007, \apj, 665, 1057 

\bibitem[Forman et al.(2017)]{Forman16} Forman, W., Churazov, E., Jones, C., et al.\ 2017, arXiv:1705.01104 

\bibitem[Graham et al.(2008)]{Graham08} Graham, J., Fabian, A.~C., \& Sanders, J.~S.\ 2008, \mnras, 386, 278 

\bibitem[\protect\citeauthoryear{Heinz, Reynolds, \& Begelman}{1998}]{1998ApJ...501..126H} Heinz S., Reynolds C.~S., Begelman M.~C., 1998, ApJ, 501, 126 

\bibitem[\protect\citeauthoryear{Hillel \& Soker}{2016}]{2016MNRAS.455.2139H} Hillel S., Soker N., 2016, MNRAS, 455, 2139 

\bibitem[Landau(1945)]{Landau45}Landau, L.D., 1945, Appl. Math. Mech., 9, 286   


\bibitem[\protect\citeauthoryear{Masuyama, Shigeyama, \& Tsuboki}{2016}]{2016PASJ...68...22M} Masuyama M., Shigeyama T., Tsuboki Y., 2016, PASJ, 68, 22 

\bibitem[\protect\citeauthoryear{McNamara et al.}{2000}]{2000ApJ...534L.135M} McNamara B.~R., et al., 2000, ApJ, 534, L135 

\bibitem[Mel'nikova(1954)]{Melnikova54} Mel'nikova, N. S., 1954, Zh. Mekhanika 3,2535



\bibitem[\protect\citeauthoryear{Randall et al.}{2011}]{2011ApJ...726...86R} Randall S.~W., et al., 2011, ApJ, 726, 86 


\bibitem[Richtmyer 
\& Morton(1967)]{R&M67} Richtmyer, R.~D., \& Morton, K.~W.\ 1967, Interscience Tracts in Pure and Applied Mathematics, New York: Interscience, 1967, 2nd ed.,  

\bibitem[Sakurai(1954)]{Sakura54} Sakurai, A.\ 1954, Journal of the Physical Society of Japan, 9, 256 


\bibitem[Sedov(1959)]{Sedov59} Sedov, L.~I.\ 1959, Similarity 
and Dimensional Methods in Mechanics, New York: Academic Press, 1959

\bibitem[Simionescu et al.(2009)]{Simionescu09} Simionescu, A., Roediger, E., Nulsen, P.~E.~J., et al.\ 2009, \aap, 495, 721 

\bibitem[Soker(2016)]{Soker16} Soker, N.\ 2016, \nar, 75, 1 


\bibitem[Tang 
\& Wang(2005)]{T&W05} Tang, S., \& Wang, Q.~D.\ 2005, \apj, 628, 205

\bibitem[Taylor(1946)]{Taylor46} Taylor, G.~I.\ 1946, 
Proceedings of the Royal Society of London Series A, 186, 273 


 
\bibitem[Truelove 
\& McKee(1999)]{TM99} Truelove, J.~K., \& McKee, C.~F.\ 1999, \apjs, 120, 299  
 


\bibitem[Xiang et al.(2009)]{Xiang09} Xiang, F., Rudometkin, E., Churazov, E., Forman, W., B{\"o}hringer, H.\ 2009, \mnras, 398, 575 

 \bibitem[\protect\citeauthoryear{Zhuravleva et al.}{2014}]{2014Natur.515...85Z} Zhuravleva I., et al., 2014, Nature, 515, 85 

\bibitem[\protect\citeauthoryear{Zhuravleva et al.}{2016}]{Zhuravleva16} Zhuravleva I., et al., 2016, MNRAS, 458, 2902 
 
 
\end{thebibliography}


\clearpage

\appendix
\section{}
\label{ap:a}
The exact history of the mass and energy injection by an AGN during an outburst is not known from observation. In our simulations, we apply the following simplified prescription. 

(i) All the mass  $M_{inj}$ is added at time $t=0$  to a central region with radius $r_{ej}(0)$, so that the density of this region is
\begin{equation}
\rho[r<r_{ej}(0)]=\frac{3M_{inj}}{4\pi r_{ej}^3(0)},
\end{equation}
where $r_{ej}(0)$ is the ejecta radius at $t=0$.

(ii) The energy is injected uniformly through this region over interval of time starting from t=0 to $t=t_b$ i.e.
\begin{equation}
\dot{e}(r,t)=\frac{3E_{inj}}{4\pi t_br_{ej}^3(t)}, \quad 0\leq t \leq t_b
\label{App:energy_rate}
\end{equation}
where $\dot{e}$ is the energy injection rate per unit mass per unit time. In simulations we typically consider the late phase evolution of an outburst, when the mass of the swept up gas is much larger than $M_{inj}$ and the results are essentially insensitive to the value of $M_{inj}$, as long as  the energy density in the ejecta, $E_{inj}/M_{inj}$, is much higher than the corresponding value of the ICM. With these parameters, the sound speed inside the ejecta is very high and eq. (\ref{App:energy_rate}) shall be a good assumption.

The adiabatic index of the injected plasma $\gamma_{ej}$ is assumed to be the same as the ICM index $\gamma_a=5/3$. While it is likely that the injected plasma is (at least partly) relativistic, suggesting $\gamma_{ej}\sim 4/3$, we are interested not in the internal structure of the ejecta, but rather in the structure of perturbations induced in the ICM by the outburst. These perturbations are not directly sensitive to the value of $\gamma_{ej}$, although the total energy $E_{inj}$ required to generate the same perturbations does depend on $\gamma_{ej}$. For instance, in a slow outburst case (see discussion in Section \ref{sec:energypartition}), the total outburst energy is related to the volume of the ejecta $E_{inj}\approx\gamma_{ej} PV/(\gamma_{ej} -1)$. As a result, a larger outburst energy $E_{inj}$ is needed for smaller $\gamma_{ej}$ in order to inflate a bubble with the same size. 

\section{}
\label{ap:b0}
In this appendix, we constrain our discussion to a uniform ambient medium and instantaneous outburst with $t_b\rightarrow 0$ for simplification. We consider an outburst with total injected energy $E_{inj}$ and total injected mass $M_{inj}$ in an ambient medium with density $\rho_a$ and pressure $P_a$ under spherical symmetry. We assume the energy and mass of the swept materials as $E_{sw}=4\pi P_aR^3/3(\gamma_a-1)$ and $M_{sw}=4\pi \rho_aR^3/3$ respectively, where $\gamma_a$ is the adiabatic index of the ambient medium and $R$ is the radius of the outburst. At early time, when $E_{sw}/E_{inj}\ll 1$ and $M_{sw}/M_{inj}\ll1$, the surrounding medium is not important. The expansion of the outburst is only determined by $E_{inj}$ and $M_{inj}$, and characterized by free expansion with $R\propto \sqrt{E_{inj}/M_{inj}}t_{age}$ according to dimensional analysis. As the age $t_{age}$ increases, the evolution of an outburst could fall into two different regimes depending on the dimensionless ratio 
\begin{eqnarray}
w&=&\frac{E_{inj}M_{sw}}{E_{sw}M_{inj}}=\frac{(\gamma_a-1)E_{inj}\rho_a}{M_{inj}P_a} \nonumber\\
&=&5\times 10^4(\gamma_a-1)\left(\frac{E_{inj}}{10^{57}erg}\right)\left(\frac{10^3\Msun}{M_{inj}}\right)\nonumber\\
&\times&\left(\frac{\rho_a}{0.1m_p cm^{-3}}\right)\left(\frac{1keVcm^{-3}}{P_a}\right)
\end{eqnarray}
where $m_p$ is the proton mass. 

In the first regime with $w\gtrsim 1$, the outburst could enter a phase with $E_{sw}/E_{inj}\ll 1$ and $M_{sw}/M_{inj}\gtrsim 1$ in which the injected mass $M_{inj}$ and the pressure $P_a$ of the ambient medium are not very important. The expansion of the outburst is only determined by $E_{inj}$ and $\rho_a$, and characterized by the relation $R\propto (E_{inj}t^2_{age}/\rho_a)^{1/5}$ according to dimensional analysis. This is simply the classic ST solution. As $t_{age}$ further increases, eventually we have $E_{sw}/E_{inj}\gtrsim 1$ and $M_{sw}/M_{inj}\gg 1$. Now the outburst starts to lose the memory of $E_{inj}$. The expansion of the outburst is only determined by $P_a$ and $\rho_a$, and characterized by the relation $R\propto \sqrt{P_a/\rho_a}$ according to dimensional analysis. This is of course the sound wave solution. So in the first regime, the outburst initially expands freely and then transits to the classic ST solution. As $t_{age}$ further increases, the strong shock eventually becomes a weak shock and asymptotically approaches a sound wave.
In the second regime with $w\lesssim 1$, the classic ST solution is no longer available for the system. Instead the outburst will enter a phase with $E_{sw}/E_{inj}\gtrsim 1$ and $M_{sw}/M_{inj}\ll 1$, which is very unstable to Rayleigh-Taylor instability. 

In most astrophysical applications like AGN feedback problem here, the evolution of the system falls into the first regime with $w\gg 1$. However in extreme condition, e.g.  supernova explosion in AGN driven bubble \citep{CCS11}, the evolution of the object could enter the second regime. 

\section{}
\label{ap:b}
In this appendix, we constrain our discussion to a uniform ambient medium for simplification. For a non-uniform medium we could replace the pressure and density in the following discussion with volume averaged values. For the AGN feedback problem of interest here, $M_{inj}$ is negligible and the intrinsic properties of an outburst are characterized by only $E_{inj}$ and $t_b$. $E_{inj}$ sets the characteristic scales for an outburst. Let's assume $R_E$ satisfy
\begin{equation}
E_{inj}=\frac{4\pi}{3}P_a R_E^3
\label{charac_R}
\end{equation}
where $P_a$ is the pressure in the ambient medium. $R_E$ depends on both the intrinsic properties of an outburst, i.e. $E_{inj}$, and the physical properties of the surrounding ambient medium, i.e. $P_a(r)$. The corresponding time scale for $R_E$ is simply the sound crossing time
 \begin{equation}
t_E=R_E\sqrt{\rho_a/\gamma_a P_a},
\label{charac_t}
\end{equation}
where $\rho_a$ is the ambient density.

For typical parameters of an outburst, we have
\begin{equation}
R_E=\left(\frac{3E_{inj}}{4\pi P_a}\right)^{1/3}=1.7 {\rm ~kpc} \left(\frac{E_{inj}}{10^{57}\rm erg}\right)^{1/3} \left( \frac{1\rm keV\cmc}{P_a}\right)^{1/3}
\end{equation}
and 
\begin{eqnarray}
t_E&=&\frac{(3E_{inj})^{1/3}\rho_a^{1/2}}{(4\pi)^{1/3}\sqrt{\gamma_a}P_a^{5/6}}= \rm 1.33\,Myr \,\mu^{1/2}\left( \frac{E_{inj}}{10^{57}\rm erg}\right)^{1/3} \nonumber\\
&&\times \left( \frac{n}{0.1\cmc} \right)^{1/2}\left(\frac{1\rm keV\cmc}{P_a} \right)^{5/6},
\label{t_E}
\end{eqnarray}
where $n$ is the number density of the surrounding medium and $\mu$ is the mean molecular weight. 
Through the paper, we assume $\mu =0.6$ for fully ionized medium with solar abundance. 

If the energy injection rate $L$ within $t_b$ is constant as assumed in this paper, i.e. $E_{inj}=Lt_b$, then for a continuous outburst, $t_E$ in the above discussion is equivalent to
\begin{eqnarray}
t_L&=&\frac{(3L)^{1/2}\rho_a^{3/4}}{\sqrt{4\pi}\gamma_a^{3/4}P_a^{5/4}}=\rm 8.6\,Myr \,\mu^{3/4}\left(\frac{L}{10^{45}\rm erg s^{-1}}\right)^{1/2} \nonumber\\
&&\times \left( \frac{n}{0.1\cmc} \right)^{3/4} \left( \frac{1\rm keV\cmc}{P_a}\right)^{5/4}.
\label{t_L}
\end{eqnarray}
and $t_L=t_E^{3/2}/t_b^{1/2}$.
It characterizes the time when energy stored in the material swept up per unit time becomes comparable with the energy injection rate $L$.
The corresponding length scale is
\begin{eqnarray}
R_L&=&\frac{(3L)^{1/2}\rho_a^{1/4}}{\sqrt{4\pi}\gamma_a^{1/4}P_a^{3/4}}=\rm 11.1\,kpc \,\mu^{1/4}\left( \frac{L}{10^{45}\rm erg s^{-1}}\right)^{1/2} \nonumber \\
&&\times \left( \frac{n}{0.1\cmc} \right)^{1/4} \left( \frac{1\rm keV\cmc}{P_a}\right)^{3/4}.
\label{r_L}
\end{eqnarray}


\section{}
\label{ap:num}
We use a one dimensional hydrodynamical code described in the Appendix B of \cite{TM99} to simulate the problem. It uses a Lagrangian finite differencing scheme with standard formulation of artificial viscosity as discussed in \cite{R&M67}. We replaced the artificial viscosity with 
\begin{equation}
q=\rho\left\lbrace c_2 \frac{\gamma+1}{4}|\Delta v|+\sqrt{c_2^2\left(\frac{\gamma+1}{4}\right)^2\Delta v^2+c_1^2c_s^2}\right\rbrace |\Delta v|
\end{equation}
to achieve better performance for weak shocks. $\Delta v$ is the velocity jump across a zone, $\rho $ is the density of the zone and $c_s$ is the sound speed in the zone \citep{Caramana98}. $c_1=0.5$ and $c_2=1$ are applied in the simulation to constrain the shock region within a few zones.  A Lagrangian CFL condition is utilized in all the simulation with
CFL number of 0.5 and the increase of the time step is required to be no more than $5\%$ between steps. Various parameters used in the simulation of an outburst in a uniform medium are presented in Table \ref{simulation}. 
We solve the hydrodynamic equations in the Lagrangian form
\begin{eqnarray}
\frac{\partial r}{\partial t}&=&v,\\
\frac{\partial v}{\partial t}&=& -4\pi r^2 \left(\frac{\partial P}{\partial M}+\rho \frac{\partial \phi}{\partial M}\right),\\
\frac{\partial e}{\partial t} &=&\frac{P}{\rho^2}\frac{\partial \rho}{\partial t} +\dot{e},\\
\frac{\partial}{\partial M}\left( \frac{4\pi r^3}{3}\right)&=&\frac{1}{\rho},\\
\end{eqnarray}
where $e$ is the internal energy per unit mass and $\phi$ is the gravitational potential. A static gravitational potential is assumed in the simulation and is derived by solving the hydrostatic equilibrium in the hot media at $t=0$, 
\begin{equation}
\frac{1}{\rho}\frac{dP}{dr}=-\frac{d\phi}{dr}.
\end{equation}

The mass injected in the simulation corresponds to a ratio $ E_{inj}\rho_a/M_{inj}P_a \sim 10^{-4}$, see detailed discussion in Appendix \ref{ap:b0}. A slight change of $M_{ej}$ could affect the density and temperature profile around the central ejecta bubble 
slightly while the outer structure of the outburst shall remain unchanged.

\begin{table}
\centering
\caption{Basic parameters for numerical simulations of an outburst in a uniform ambient medium}
\begin{tabular}{lc}
\hline\hline
injected mass $M_{ej}$& $1000 \Msun$ \\
outburst energy $E_{inj}$&$10^{57} \,\rm erg$\\
ambient density $n_a$& $0.1 \,\rm cm^{-3}$\\
ambient pressure $P_a$& $1 \,\rm   keVcm^{-3}$\\
\hline\hline
\end{tabular} 
\label{simulation}
\end{table}

\section{}{\label{App:shock_evolution}}
\label{ap:d}

In this Appendix, we discuss the shock evolution in an instantaneous outburst and a continuous outburst. For simplification, we assume a uniform ambient medium.

In \cite{T&W05}, it has been shown that in an instantaneous outburst the shock velocity $V_{s}$ could be reproduced within 3 percent accuracy by simply connecting the ST solution velocity and the sound speed in the ambient medium as follows  
\begin{equation}
V_{s}(t)=\left(V_{ST}^{5/3}(t)+c_s^{5/3}\right)^{3/5}=c_s\left(1+\eta\frac{t_E}{t}\right)^{3/5},
\label{velocity_fast}
\end{equation}
where $V_{ST}$ is the velocity from the ST solution, $c_s$ is the sound speed in the ambient medium and $\eta=(2\xi/5)^{5/3}(4\pi/3\gamma_a)^{1/3}\approx 0.374$ for $\gamma_a=5/3$ and $\xi=1.152$. At $t=t_E$, we have $M\approx 1.2$ which implies $t_E$ characterizing the transition from strong shock to weak shock. 

We can rewrite eq \ref{velocity_fast} to express the shock age as a function of Mach number, i.e.
\begin{equation}
t=\frac{\eta t_E}{M^{5/3}-1}.
\end{equation} 
The above equation can be used to estimate the age of an instantaneous outburst based on the known Mach number.

The radius of the shock front $R_s$ could be derived by integrating the velocity $V_{s}$ and after some calculation we obtain
\begin{equation}
R_{s}(t)=\frac{5}{2}c_s\eta t_E \left(\frac{t}{\eta t_E}\right)^{2/5} F(-\frac{3}{5},\frac{2}{5};\frac{7}{5};-\frac{t}{\eta t_E})
\label{radius_fast}
\end{equation} 
which is equivalent to eq (4) in \cite{T&W05} and $F$ is the generalized hypergeometric function.

The dynamical evolution of a continuous outburst with a constant energy injection rate $L$ and negligible ambient pressure can be described by a generalized ST solution \citep{Dokuchaev02}, in which the shock radius and velocity follow 
\begin{equation}
R_{GST}=\xi_L\left(\frac{ Lt^3}{\rho}\right)^{1/5}\quad  \mbox{ and } \quad V_{GST}=\frac{dR}{dt}=\frac{3\xi_L}{5}\left(\frac{ L}{\rho t^2}\right)^{1/5}
\end{equation} 
respectively. $\xi_L$ is a dimensionless constant and for $\gamma_a=5/3$ \cite{Dokuchaev02} found $\xi_L=0.929$. Although the generalized ST solution provided in \cite{Dokuchaev02} requires a special distribution of energy injection in space which is different from our numerical set up, we find the solution provides a good fit to the simulation results for the time range investigated in this paper.

Following the idea in \cite{T&W05} for an instantaneous outburst, we find the velocity of a continuous outburst with constant energy injection rate $L$ can be approximated by connecting the generalized ST solution velocity and the sound speed together as follows
\begin{equation}
V_{s}(t)=\left(V_{GST}^{5/2}(t)+c_s^{5/2}\right)^{2/5}=c_s\left(1+\eta_L \frac{t_L}{t} \right)^{2/5}
\label{velocity_slow}
\end{equation}
where $\eta_L=(3\xi_L/5)^{5/2}(4\pi/3\gamma_a)^{1/2}\approx 0.368$ for $\gamma_a=5/3$.  We can rewrite eq. \ref{velocity_slow} to express the shock age as a function of Mach number, i.e.
\begin{equation}
t=\frac{\eta_L t_L}{M^{5/2}-1}.
\end{equation} 
The above equation can be used to estimate the age of a continuous outburst at arbitrary Mach number $M$. 

The corresponding shock radius can be obtained by integrating eq. (\ref{velocity_slow}) over time. After some calculation we get
\begin{equation}
R_{s}(t)=\frac{5}{3}c_s\eta_L t_L \left(\frac{t}{\eta_L t_L}\right)^{3/5} F(-\frac{2}{5},\frac{3}{5};\frac{8}{5};-\frac{t}{\eta_L t_L})
\label{radius_slow}
\end{equation} 
which can reproduce the simulation results within a few percent. $F$ again is the generalized hypergeometric function.

For a continuous outburst, at late times when $t_{age}\gtrsim t_E$, the energy left in ejecta is  $E_{ej}=Lt_{age}/\gamma_{ej}=PV/(\gamma_{ej}-1)$. The corresponding ejecta radius then becomes
\begin{equation}
R_{ej}(t)=\left[\frac{3(\gamma_{ej}-1)Lt}{4\pi \gamma_{ej} P}\right]^{1/3}=R_L\left[\frac{3(\gamma_{ej}-1)t}{4\pi \gamma_{ej} t_L}\right]^{1/3}.
\label{radius_ejecta}
\end{equation}
According to eq (\ref{radius_slow}) and (\ref{radius_ejecta}), the $R_s/R_{ej}$ ratio for a continuous outburst at $t_{age}\gtrsim t_E$ simply satisfies
\begin{eqnarray}
\frac{R_s(t)}{R_{ej}(t)}&=&\xi_L\left(\frac{4\pi}{3\gamma_a} \right)^{1/5} \left[\frac{\gamma_{ej}}{(\gamma_{ej}-1)}\right]^{1/3}\nonumber \\
&\times &\left(\frac{t}{t_L}\right)^{4/15}F(-\frac{2}{5},\frac{3}{5};\frac{8}{5};-\frac{t}{\eta_L t_L})\nonumber \\
\end{eqnarray}
where $\eta_L=(3\xi_L/5)^{5/2}(4\pi/3\gamma_a)^{1/2}\approx 0.368$ and $\xi_L=0.929$ for $\gamma_a=5/3$ \citep{Dokuchaev02}. $F$ again is the generalized hypergeometric function.

\begin{figure}
 \begin{center}
 \includegraphics[width=\columnwidth]{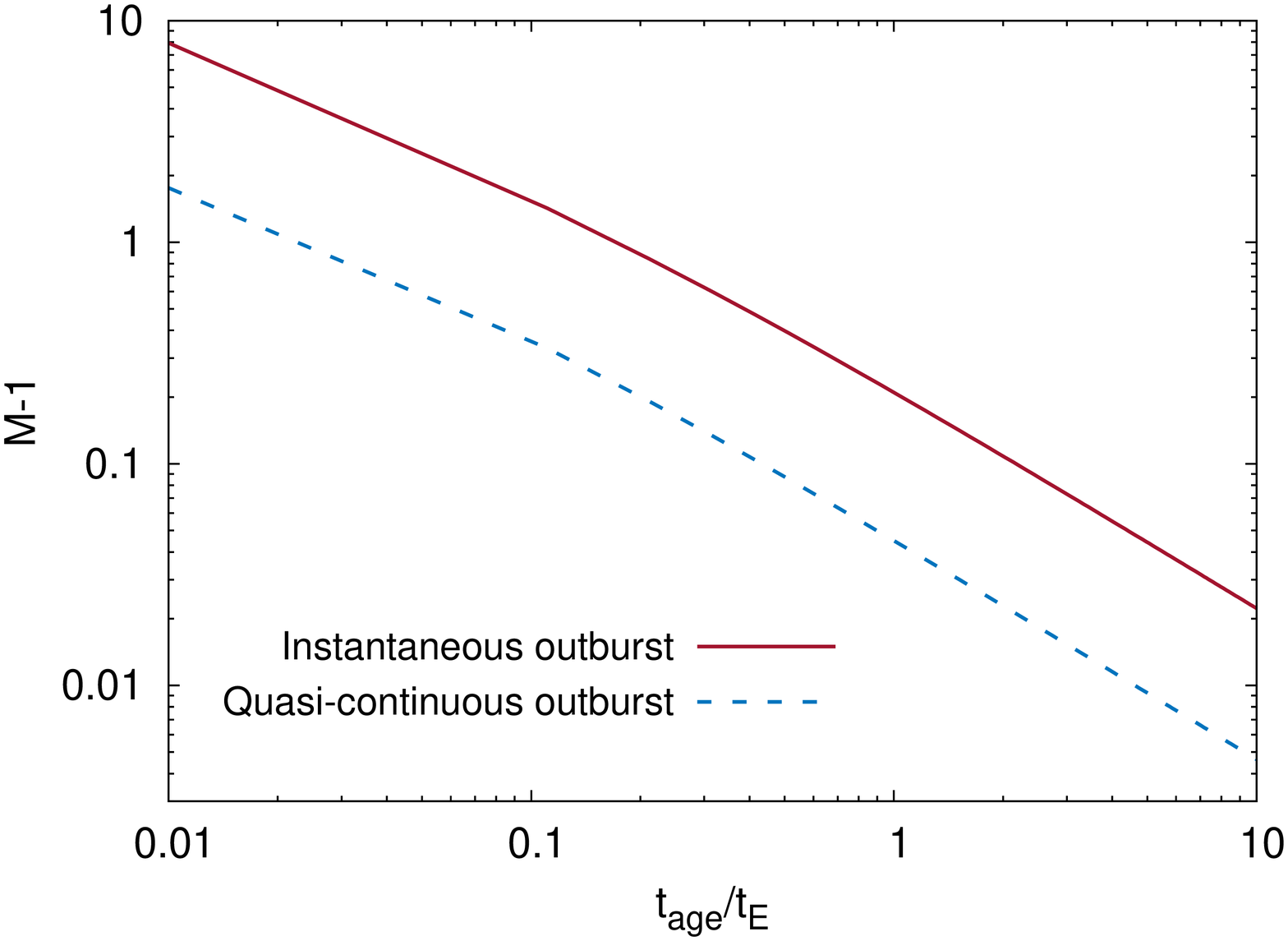} 
  \caption{M-1 as a function of $t_{age}/t_E$ according to analytical approximation eq. \ref{velocity_fast} and \ref{velocity_slow}. Red solid line is for an instantaneous outburst and the blue dashed line is for a quasi-continuous outburst with $t_b=10t_E$.} 
    \label{Machnumber}
 \end{center}
 \end{figure}

Fig. \ref{Machnumber} illustrates how the Mach number decrease with time for both an instantaneous outburst and a quasi-continuous outburst with $t_b/t_E=10$ based on analytical approximations derived in this Appendix. As $t_b/t_E$ gradually increases, i.e. longer outbursts, the value of M-1 will decrease even faster accordingly. Thus a quasi-continuous outburst with $t_b/t_E\gg 1$ is characterized by a weak shock propagating at essentially the sound speed over almost the entire duration of the outburst.

\section{}{\label{App: M87_Perseus}}
The unperturbed density and temperature distribution adopted in our simulation for M87 and the Perseus cluster are presented as below. 

In M87 \citep{Xiang09}
\begin{equation}
n_e(r)=0.22\,{\rm cm^{-3}}\left[1+\left(\frac{r}{r_c}\right)^2\right]^{-3\beta/2}
\end{equation}
where $\beta=0.33$ and $r_c=0.93 \rm kpc$ and
\begin{equation}
T_e(r)=1.55\,{\rm keV}\left[1+\left(\frac{r}{10.23 {\rm kpc}}\right)^2 \right]^{0.18}.
\end{equation}
The total number density to electron number density ratio, i.e. $n_t/n_e$, is assumed to be $1.9$ through the paper.
A static gravitational potential is assumed in the simulation and solved by requiring that the ambient medium satisfy hydrostatic equilibrium.

For the Perseus cluster, we adopted a slightly modified version of the profiles from \cite{Churazov03}, re-scaled  to a Hubble constant of 70km/s/Mpc.
\begin{equation}
n_e(r)=\frac{5.36\times 10^{-2} {\rm cm^{-3}}}{[1+(r/55{\rm kpc})^2]^{1.8}}+\frac{4.6\times 10^{-3}{\rm cm^{-3}}}{[1+(r/ {\rm 200kpc})^2]^{0.87}}
\end{equation}
and
\begin{equation}
T_e(r)=9.0 {\rm keV} \frac{1+(r/60 {\rm kpc})^{2.5}}{2.7+(r/60{\rm kpc})^{2.6}}.
\end{equation}

\begin{figure}
 \begin{center}
 \includegraphics[width=\columnwidth]{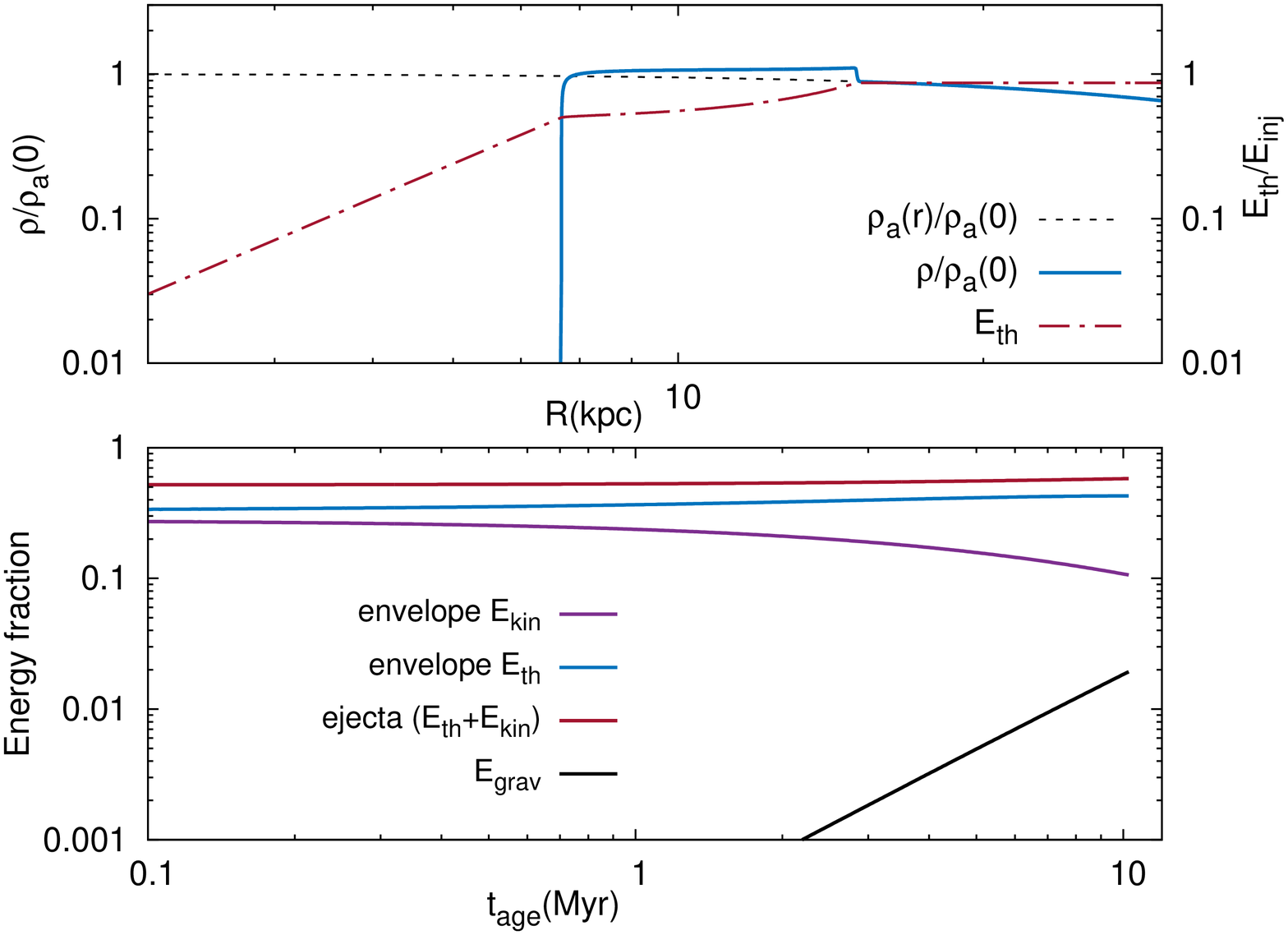} 
  \caption{Numerical simulation for an outburst in the Perseus cluster with total injected energy $E_{inj}\approx 1.1\times 10^{59}\rm erg$ energy and $t_b=t_{age}=10.3\rm Myrs$. The upper panel shows the normalized density distribution (red solid) and the accumulative thermal energy distribution $E_{th}$ (dotted blue). The dashed green line is the normalized unperturbed density distribution. The lower panel presents the time evolution of the different energy components during the expansion of an outburst.} 
    \label{1e59}
 \end{center}
 \end{figure}

\begin{figure}
 \begin{center}
 \includegraphics[width=\columnwidth]{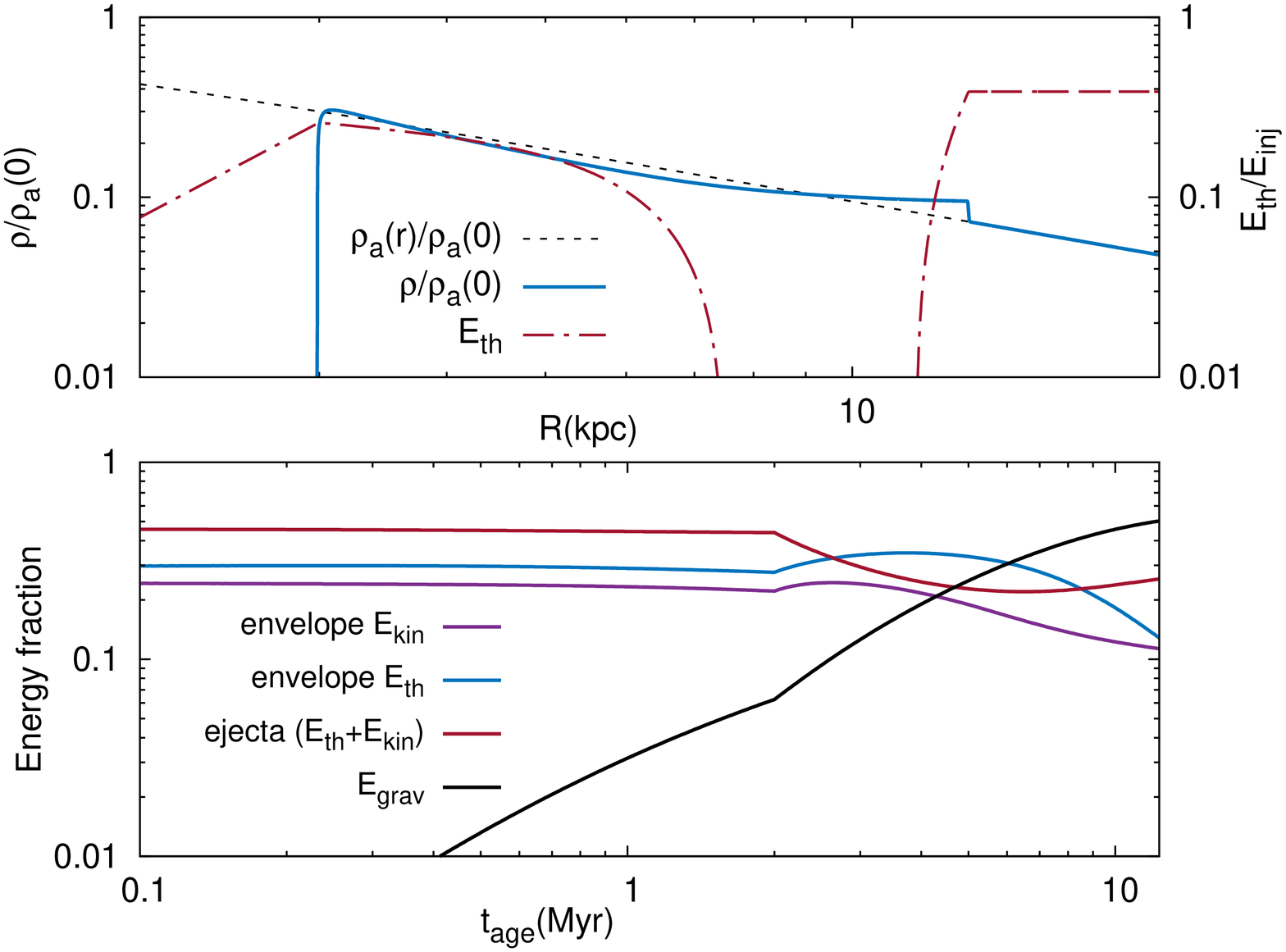} 
  \caption{Numerical simulation for an outburst in M87 with total injected energy $E_{inj}=5.5\times 10^{57}\rm erg$, outburst duration time $t_b=2\rm Myrs$ and age $t_{age}=12.3Myr$. The upper panel shows the normalized density distribution (blue solid) and the accumulative thermal energy distribution $E_{th}$ (red dotted dashed). The dotted black line is the normalized unperturbed density distribution. The lower panel presents the time evolution of different energy components during the expansion of an outburst.} 
    \label{M87}
 \end{center}
 \end{figure}

The best fit simulation results for the Perseus cluster and M87 are presented in Fig. \ref{1e59} and \ref{M87} respectively, which are consistent with previous discussion in \cite{Zhuravleva16}and \cite{Forman16}. According to the above simulations, it is interesting to see that the Perseus cluster seems to have ongoing energy injection for quite a long time until now, while the AGN activity in M87 seems to be quenched a long time ago. As a result, the outbursts in the Perseus cluster and M87 show quite different dynamical structures.  

The continuous outburst in the Perseus cluster mainly consists two parts, a hot ejecta bubble and a WSH gas envelope, which are the same as that in a uniform medium (see Fig. \ref{piston_region}). The WSH envelope in the Perseus cluster has physical properties close to the unperturbed ambient medium except the region close to the shock front, which is also similar to outbursts in a uniform medium. A significant amount of energy, more than $50\%$ of $E_{inj}$, is stored in the central ejecta bubble, which is consistent with a continuous outburst in a uniform medium. No WLS appears in the WSH envelope as the energy injection driving such an outburst hasn't quenched yet. Only about $2\%$ of injected energy $E_{inj}$ is transferred to the gravitational energy, which indicates the surrounding medium swept up by the shock front doesn't deviate too much from a uniform distribution. 

The simulation results for M87 are presented in Fig. \ref{M87} and are consistent with that from an intermediate outburst in a uniform medium. The spatial structure of the outburst also contains two parts, a central ejecta bubble and a WSH gas envelope. But the WSH envelope in M87 is dominated by the WLS while the structure between $R_{env}$ and $R_w$ (see Fig. \ref{various_region} for example), which has physical properties close to the unperturbed medium, is negligible in the figure. It implies that the $t_{age}/t_E$ ratio in M87 is not very large and the WLS structure has just detached from the central ejecta bubble.  The energy stored in the ejecta bubble of M87 is smaller than the Perseus cluster and is about $30\%$ which is consistent with an intermediate outburst . In the mean time, about $50\%$ of the injected energy $E_{inj}$ is transferred to the gravitational energy due to the steep density profile in the core of M87. As a result, the time evolution of various energy components in M87 becomes more complicated than that in a uniform medium as the surrounding environment now has a strong effect in shaping the dynamical structure of the outburst. 

Overall the spatial structure of an outburst in a non-uniform medium is qualitatively similar to that in the uniform medium. 

\section{}{\label{App:PVM_prove}}
In this Appendix, we investigate the validity of Fig. \ref{PVM} and show that the contours in the figure derived with simulation in a uniform medium are still a good approximation for AGN driven outbursts in galaxies and clusters with non-uniform media. In a non-uniform medium, the effective characteristic time $t_E$ depends on the spatial structure of the ambient medium and evolves with time during the evolution. As a result, different clusters or galaxies with different density and pressure profile exhibit different foot prints in Fig. \ref{PVM}. Therefore it is impossible to make a complete comparison between the uniform medium case and the non-uniform medium case. We instead use M87 and the Perseus cluster as examples to illustrate possible difference between the uniform medium and non-uniform medium simulation results. As we will show later, for galaxy clusters with shallow density and pressure profile (power law index $\lesssim$2) like M87 and the Perseus cluster, the deviation from a uniform medium results is not very significant and Fig. \ref{PVM} can still be considered as a good approximation for AGN driven outbursts in galaxies and clusters with non-uniform media.

\begin{figure}
 \begin{center}
 \includegraphics[width=\columnwidth]{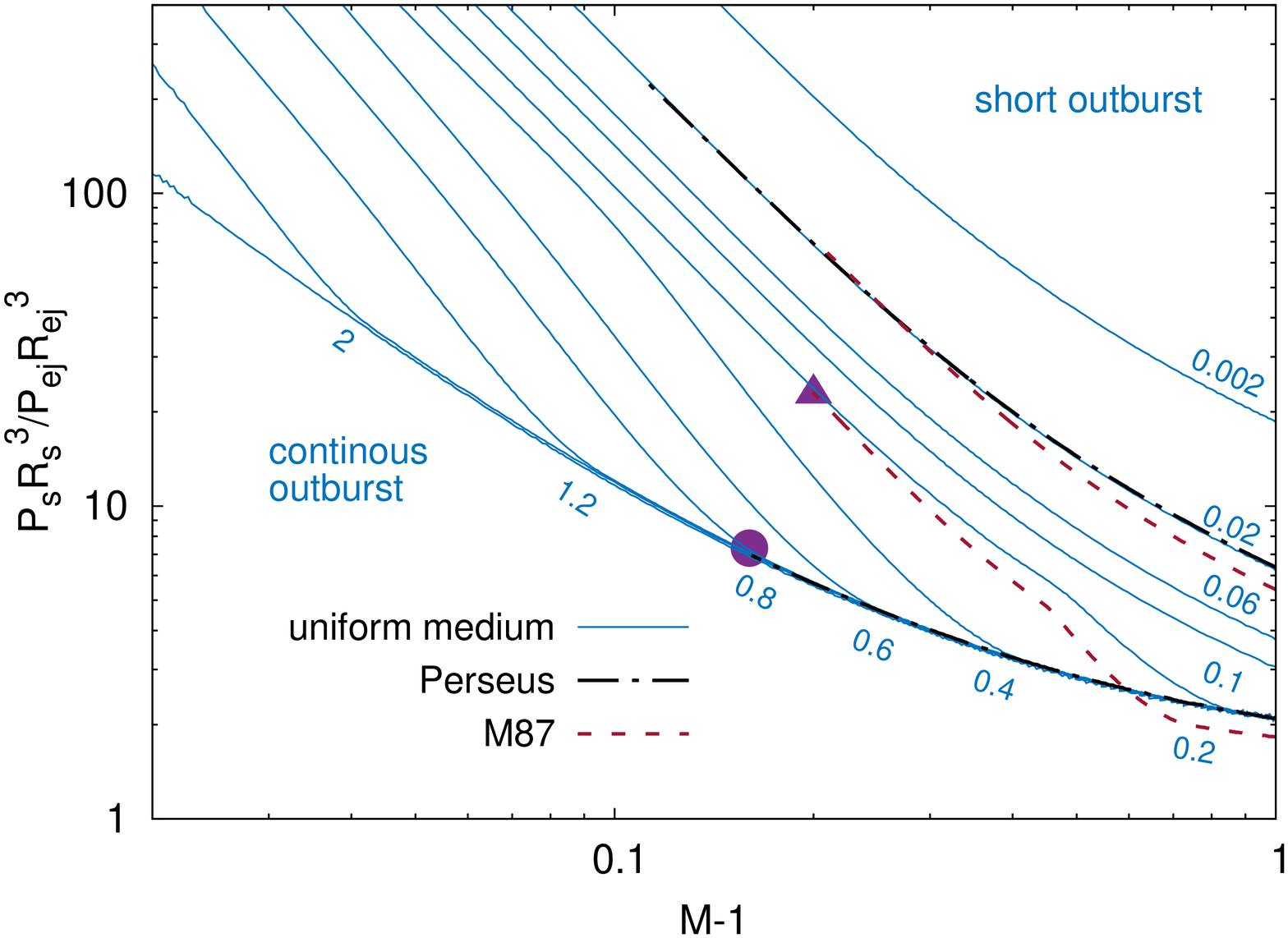} 
  \caption{x axis: $M-1$ where $M$ is the Mach number of the shock wave; y axis: $G=P_sR^3_s/P_{ej}R_{ej}^3$ where $R_s$ is the shock radius, $R_{ej}$ is the ejecta radius, $P_{ej}$ is the pressure at $R_{ej}$ and $P_s$ is the pressure at $R_s$. Blue solid lines represent the contours with different $t_b/t_E$ derived in a uniform medium. Blue numbers mark the corresponding $t_b/t_E$ ratio of the line. The bottom blue solid line can be considered as a representative of a continuous outburst. Black dotted-dashed lines are for the Perseus cluster while red dashed lines are for M87. For both the Perseus cluster and M87, the upper curves presented are short outbursts with $t_b/t_E=0.02$ ($t_E$ is calculated from the density and pressure in the galaxy center) and the lower curves show our best fit numerical models as examples for long outbursts.}
    \label{PVM_prove}
 \end{center}
 \end{figure}

In Fig. \ref{PVM_prove}, we present the contours of $t_b/t_E$ (blue solid) as a function of $G=P_sR_s^3/P_{ej}R_{ej}^3$ and $M-1$ based on simulation in a uniform medium. We also plot two curves, one for a short outburst and the other for a long outburst, with both the Perseus cluster (black dotted-dashed) and M87 (red dashed) profiles to compare the non-uniform medium results with uniform medium ones. The short outburst is simulated with $t_b/t_E=0.02$, where $t_E$ is calculated with pressure and density in the center of the galaxy. The long outburst chosen for illustration is our best fit numerical model for the two objects shown in Table \ref{simulation_nonuniform}.

The black dotted-dashed curves with density and pressure profile of the Perseus cluster agree with the corresponding uniform medium lines very well. It is mainly because the core of Perseus cluster has very flat density and pressure profile. Moreover, for the time and Mach number range we are interested in the outburst mainly expands in the flat core with an almost uniform medium and hasn't reached the steep envelope of the galaxy yet. The red dashed curves in M87 profiles show larger deviation from the uniform medium lines. At early times with $M\sim 2$, we found the curves for M87 fall below the corresponding uniform medium line. This is mainly because the temperature profile in M87 increases with radius instead of decreasing with radius which is different from the Perseus cluster. Because of this, the pressure profile in M87 exhibits a broken power law distribution with a steeper power in the inner part and a shallower power law in the outer part. The break happens at around $10 \rm kpc$. When the shock front is in the shallower power law region while the ejecta bubble is in the steep power law region, the $P_sR_s^3/P_{ej}R^3_{ej}$ ratio can decrease with time and become smaller than the uniform medium results. The two simulations of M87 are stopped with shock radius about 15kpc which indicates the outburst indeed pass through this region during the expansion. As time increases and M decreases, the curves start to bend up and approach the uniform medium line with smaller $t_b/t_E$ ratios. It is because the outburst enters the envelope of M87 with a steeper profile. The pressure and density are decreasing with radius very quickly in the envelope. As a result, the characteristic time $t_E$ is increasing with time and the effective $t_b/t_E$ of the outburst is decreasing with time which cause the curves to bend up. 

Although the curves for the Perseus cluster and M87 behave slightly differently due to the different density and temperature distributions. For the Mach number range $1.1-1.3$ and $t_b/t_E$ ratio between $0.01$ and $ 1$, which we are particularly interested in, the deviation is not very significant. Thus the contours derived in simulation with a uniform medium are still a good approximation for AGN driven outbursts in galaxies and clusters with shallow density and pressure profiles (power law index $\lesssim 2$).

\bsp	
\label{lastpage}
\end{document}